\documentclass[aps,prl,times,twocolumn,superscriptaddress,showpacs,a4paper]{revtex4-1}

\usepackage{colortbl}
\usepackage{amsfonts,amsmath,amssymb}
\usepackage[dvips]{graphicx}
\usepackage{hyperref}
\newcommand{\ie}{{\it i.e. }}

\renewcommand{\=}{\,=\,}

\renewcommand{\-}{\,-\,}

\begin{document}


\title{Multiferroicity in spin ice: towards a magnetic crystallography of Tb$_{2}$Ti$_{2}$O$_{7}$ in a field}

\author{L.D.C Jaubert}
\affiliation{Okinawa Institute of Science and Technology Graduate University,
Onna-son, Okinawa 904-0395, Japan}

\author{R. Moessner}
\affiliation{Max-Planck-Institut f\"ur Physik komplexer Systeme, 01187 Dresden, Germany}

\date{\today}
\begin{abstract}
We combine two aspects of magnetic frustration, multiferroicity and emergent quasi-particles in spin liquids, by studying magneto-electric monopoles. Spin ice offers to couple these emergent topological defects to external fields, and to each other, in unusual ways, making possible to lift the degeneracy underpinning the spin liquid and to potentially stabilize novel forms of charge crystals, opening the path to a ``magnetic crystallography''. In developing the general phase diagram including nearest-neighbour coupling, Zeeman energy, electric and magnetic dipolar interactions, we uncover the emergence of a bi-layered crystal of \textit{singly-charged} monopoles, whose stability, remarkably, is strengthened by an external [110] magnetic field. Our theory is able to account for the ordering process of Tb$_{2}$Ti$_{2}$O$_{7}$ in large field for reasonably small electric energy scales.
\end{abstract}
\maketitle

 
By providing mechanisms for strong magneto-electric coupling, frustration has become a key ingredient in multiferroics~\cite{Kimura03a,Hur04a,Katsura05a,Sergienko06a,Mostovoy06a,Khomskii06a,Cheong07a,Tokura14a}. While the search for high-temperature multiferroics is appealing for technological application such as memory devices~\cite{Tokura14a}, frustration opens a window on novel fundamental properties of magnetic matter at low temperature where even weak perturbations can play an important role. This holds especially for the collective behaviour of spin liquids in a wide range of compounds from rare-earth~\cite{Gardner10a} and copper~\cite{Shores05a,Okamoto09a,Quilliam11b} oxides to organic Mott insulators~\cite{Kanoda11a} or iridates~\cite{Modic14a,Takayama14a}.

In spin ice materials, the constraints imposed by frustration support an extensively degenerate ground state where magnetic fluxes are locally conserved~\cite{Harris97a}. Such flux conservation can be described as a divergence-free condition, categorizing the spin ice ground state as a Coulomb spin liquid by analogy with Maxwell's electromagnetism~\cite{Isakov05a, Fennell09a,Henley10a}, where excitations take the form of classical magnetic monopoles (Fig.~\ref{fig:latt}.$c$-$d$)~\cite{Castelnovo08a}.

In addition to their magnetic properties, it has been recently theorized that magnetic monopoles could also carry an electric dipole moment~\cite{Khomskii12a} (Fig.~\ref{fig:latt}.$c$). 
Here we shall investigate the multiple facets of such magneto-electric coupling, as an unexplored generic ordering process in rare-earth pyrochlores, able to lift the degeneracy of spin liquids and to manipulate topological excitations in frustrated magnets. Our results are double. First of all, we give a precise description of the mosaic of competing phases in our multiferroic spin ice model (Eq.(\ref{eq:ham})). We show how a \textit{ferromagnetic double-layer structure of monopoles} (DL) is stabilized by electric dipolar interactions and even enhanced by a magnetic field in the [110] direction. We then use this double-layer structure as a signature of multiferroicity in rare-earth oxides able to account for recent experiments on the spin liquid candidate Tb$_{2}$Ti$_{2}$O$_{7}$ in a field.\\


\paragraph{The model --}we consider classical Ising spins $\vec S$ aligned with their local easy-axes on the pyrochlore lattice supporting electric moments $\vec P$ induced by magneto-electric coupling~\cite{Khomskii12a} (Fig.~\ref{fig:latt}), interacting via nearest neighbour spin coupling, Zeeman energy, magnetic and electric dipolar interactions
\begin{eqnarray}
\label{eq:ham}
\mathcal{H}&=& J\;\sum_{\langle ij\rangle}\; \vec S_{i}\cdot \vec S_{j} \;-\; \vec h \cdot \sum_{i}\;\vec S_{i}\\\nonumber
&+& D_{m}r_{m}^{3}\;\sum_{i>j} \frac{\vec S_{i}\cdot \vec S_{j} \, -\, 3 \left(\vec S_{i}\cdot \vec e_{ij}\right)\left(\vec S_{j}\cdot \vec e_{ij}\right)}{r_{ij}^{3}}\\\nonumber
&+& D_{e}r_{e}^{3}\;\sum_{\alpha>\beta} \frac{\vec P_{\alpha}\cdot \vec P_{\beta} \, -\, 3 \left(\vec P_{\alpha}\cdot \vec e_{\alpha\beta}\right)\left(\vec P_{\beta}\cdot \vec e_{\alpha\beta}\right)}{r_{\alpha\beta}^{3}}\nonumber
\end{eqnarray}
%
\begin{figure}[h]
\centering\includegraphics[width=9cm]{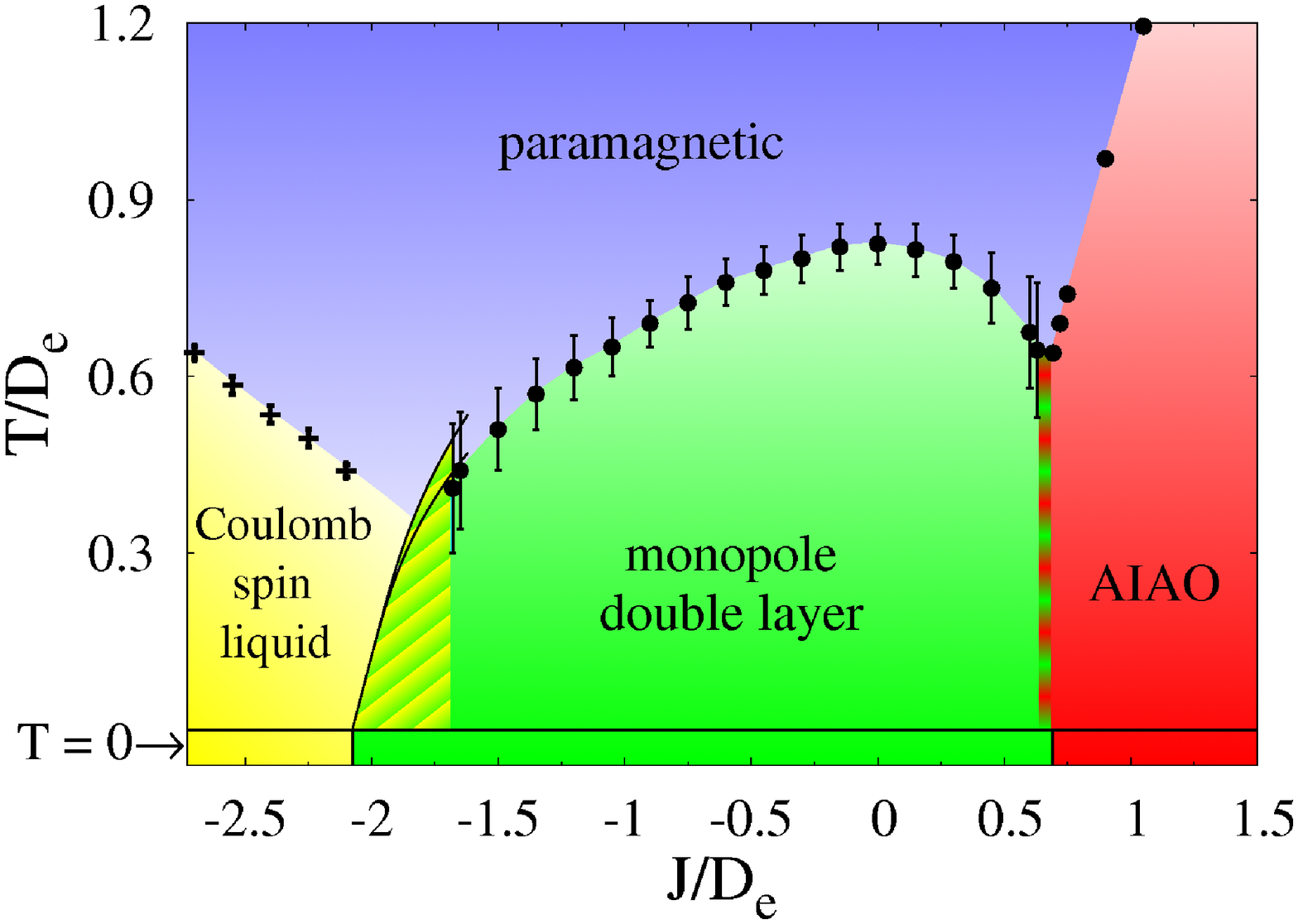}
\caption{When $D_{m}=h=0$, the electric dipoles stabilize a monopole double layer (DL - green, see Fig.~\ref{fig:latt}.$a$), in competition with all in / all out order (AIAO - red), and the Coulomb spin liquid (yellow). The circles (crosses) are the transition (crossover) temperatures obtained from Monte Carlo simulations. In the hatched regions, even if $T=0$ calculations confirm the energetic stability of the double-layer structure, the first order nature of the transition prevents full thermalization of the simulations. The solid lines are upper and lower mean field estimates of the boundary. See Appendix for details on simulations and calculations.}
\label{fig:PD1}
\end{figure}

where $i,j$ and $\alpha,\beta$ are respectively indices for magnetic spins on the pyrochlore lattice and electric dipoles on the diamond lattice. $r_{m}$ and $r_{e}=\sqrt{3/2}\;r_{m}$ are the respective nearest neighbour distances. The nearest neighbour vector $\vec e_{ij}$, magnetic $\vec S_{i}$ and electric $\vec P_{\alpha}$ moments have unit length. The size of the moments $\mu$ and $p$ is included in the energy-scale prefactors
\begin{eqnarray}
D_{m}=\frac{\mu_{0}\;\mu^{2}}{4\pi\,r_{m}^{3}},\quad
D_{e}=\frac{p^{2}}{4\pi\varepsilon_{0}\,r_{e}^{3}},\quad
\vec h = \mu_{0}\;\mu\;\vec H
\label{eq:NRJscale}
\end{eqnarray}
where $\mu_{0}$ and $\varepsilon_{0}$ are respectively the vacuum magnetic permeability and electric permittivity and $\vec H$ is the external magnetic field.

The Hamiltonian is studied via classical Monte Carlo simulations,  using parallel tempering, worm and single-spin-flip Metropolis algorithm. The dipolar energies have been computed with the Ewald summation~\cite{Deleeuw80a,Hertog00a}, in absence of demagnetization factor in order to develop a sample-independent theory~\cite{Melko04a}. All spin configurations are given in the Appendix.\\

\begin{figure}
\centering\includegraphics[width=\columnwidth]{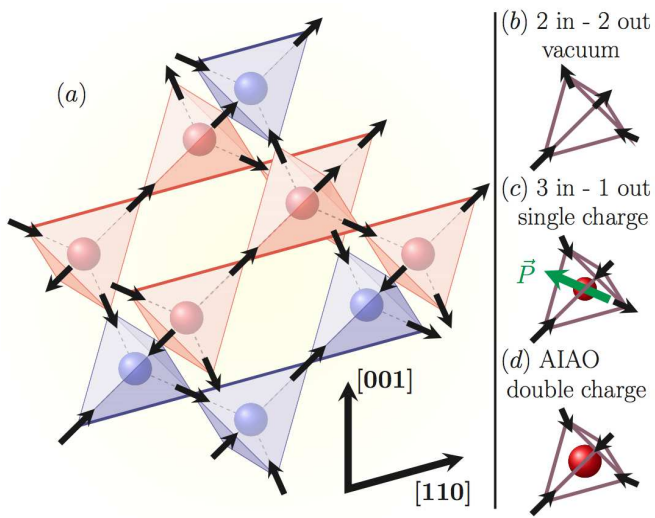}
\caption{($a$) Electrically induced ground state of our multiferroic spin ice model, made of alternative bi-layers of positive (blue) and negative (red) magnetic charges stacked along a [001] axis. The $\alpha-$chains (indicated by thick bonds) carry a saturated [110] magnetisation. ($b-d$) There are three different kinds of configurations for a given tetrahedron: 2 in - 2 out (vacuum of charge forming the Coulomb spin liquid), 3 in - 1 out (single magnetic charges carrying an electric moment $\vec P$ whose direction is dictated by the minority (here outward) spin and independent of the sign of the magnetic charge because of time-reversal symmetry, forming the double-layer structure), and 4 in (double charges forming the AIAO order).
}
\label{fig:latt}
\end{figure}

\paragraph{Monopole Double Layer --\hspace{-0.2cm}}
first of all, what happens in absence of magnetic interactions, \textit{i.e.} $D_{m}=h=J=0$ ? We find that electric dipoles induce a bi-layer structure of single charges with zero polarization and saturated magnetization along the [110] axis (Fig.~\ref{fig:latt}.$a$). Because the electric field is even under time reversal, the apparition of such magnetization has to be spontaneous. This configuration is unfrustrated at the nearest-neighbour level, whose contribution constitutes 96\% of the total energy. To our knowledge, magneto-electric coupling is the first \textit{intrinsic} mechanism favouring single magnetic charges down to zero temperature.\\


\paragraph{Local chemical potential $J$ --\hspace{-0.2cm}}
in terms of monopoles, $J$ plays the role of a chemical potential favoring the Coulomb spin liquid for $J<0$ (vacuum of charge) and the AIAO double-charge crystal for $J>0$~\cite{Moessner98b,Bramwell98a}, making single charges gapped topological excitations in both cases. However, the previously observed $J=0$ double-layer structure turns out to be robust over a large range of values for $J/D_{e}\in[-2.08:0.69]$ (Fig.~\ref{fig:PD1}), raising the question on the nature of the mechanism able to stabilize such ``excitations''.

For instance when $J<0$, creating a pair of single charges out of the Coulomb spin liquid costs $|4J/3|$ while the energy gain is at most $-2D_{e}/3$, making such monopole-pair creation unfavorable for $J/D_{e}<-1/2$. An energetically stable cluster of bi-layered monopoles thus needs to get bigger and bigger as $J$ decreases in order to minimize its surface-over-volume ratio. The need for this kind of nucleation process to seed and grow a cluster makes the transition first order and prevents full thermalization of the simulations in the vicinity of the extensively degenerate Coulomb spin liquid (see yellow/green hatched region in Fig.~\ref{fig:PD1}). As a consequence, for $J\approx -2 D_{e}$, an experimental cooling down protocol would probably fall out-of-equilibrium; once the magnet enters the Coulomb spin liquid with a low density of monopoles, it will be difficult to nucleate a big enough cluster of magnetic charges to crystallize the double-layer structure. Such phenomena also exist for $J>0$, but to a lesser extent because of the low entropy of the AIAO ordered phase (see red/green hatched region in Fig.~\ref{fig:PD1}).

Hence, the magneto-electric opportunity to stabilize monopole excitations comes at the cost of large (free) energy barriers and multiple metastable states, which can naturally account for strong out-of-equilibrium effects in pyrochlores. In order to build a comprehensive and experimentally relevant picture of the problem, let us now include magnetic dipolar interactions, before adding a magnetic field able to tune these energy barriers, and finally applying our theory to experiments.\\


\paragraph{Long range magnetic dipolar interactions $D_{m}$ --\hspace{-0.2cm}} since the electric polarization is coming from magneto-electric coupling, it would be improper to neglect the $D_{m}$ energy scale, especially if we keep in mind rare-earth materials with potentially large magnetic moments. In spin ice, magnetic dipolar interactions are responsible for the effective Coulomb interactions between monopoles~\cite{Castelnovo08a}. The property of ``projective equivalence''~\cite{Isakov05a} ensures the quasi-degeneracy of the Coulomb spin liquid in presence of magnetic dipolar interactions~\cite{Hertog00a}, which is only weakly lifted in favour of the 2 in - 2 out long range ordered dipolar spin ice (ODSI) state for $T\ll D_{m}$~\cite{Siddharthan99a,Melko01a}.

\begin{figure}[ht]
\centering\includegraphics[width=7cm]{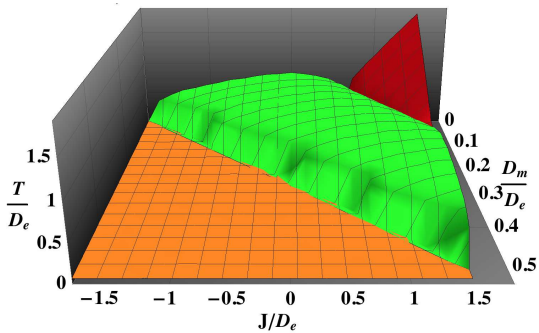}
\caption{Phase diagram of $J/D_{e}$ and $D_{m}/D_{e}$ obtained from Monte Carlo simulations. The vertical axis is the normalized transition temperature into the AIAO (red), double-layer (green) or ODSI (orange) phase. See Appendix for details.}
\label{fig:PD2}
\end{figure}

This is why, by favoring 2 in - 2 out configurations, opposing the proximity of same-charge monopoles and hindering long-range ferromagnetic order, the $D_{m}$ interaction seems very unfavorable to the DL phase. Indeed for negative $J$, the double-layer phase makes way for the ODSI at finite $D_{m}$ (Fig.~\ref{fig:PD2}). Remarkably, the projective equivalence, valid for magnetic dipolar interactions in the Coulomb spin liquid but not electric ones in the DL phase, makes the transition temperature one order of magnitude smaller from the DL to the ODSI phase.

However the nearest neighbour contribution of the $D_{m}$ term decreases the monopole chemical potential which is four times bigger for double charges than for single ones~\cite{Castelnovo08a}. The counter-intuitive consequence is that magnetic dipolar interactions favour DL order for positive values of $J$ where it was absent at $D_{m}=0$. Because magnetic Coulomb interactions in spin ice are four orders of magnitude smaller than between bare electric charges at the same distance, a relatively low-energy coupling such as $D_{e}$ is sufficient to counter-balance the repulsion between neighbouring magnetic charges, opening the path for \textit{novel crystal structures made of magnetic monopoles}.

\begin{figure}[ht]
\centering\includegraphics[width=7cm]{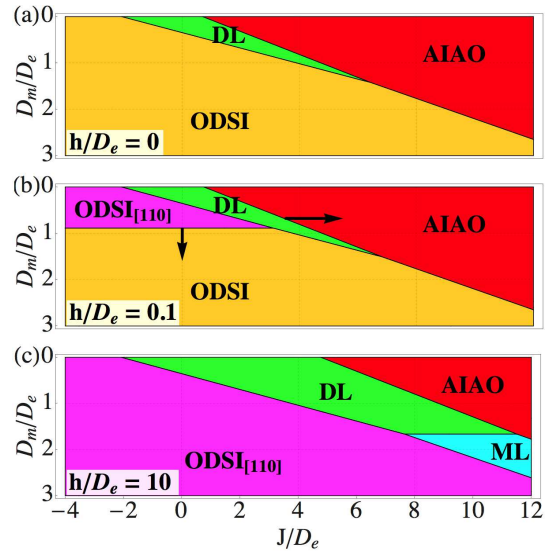}
\caption{Zero temperature phase diagram in a [110] field $h$ obtained from Ewald summation. The arrows show the evolution of the boundaries as $h$ increases. The DL phase is further stabilized by $h$ over the AIAO order, but restricted to small values of $D_{m}$ by the apparition of a monolayer phase of monopoles (ML, in cyan).}
\label{fig:PDh}
\end{figure}

\paragraph{A [110] magnetic field $h$ --\hspace{-0.2cm}} in the absence of magneto-electric couplings, such a field polarizes half of the spins along the $\alpha-$chains (Fig.~\ref{fig:latt}.$a$) while magnetic dipolar interactions align the remaining spins along antiferromagnetically-ordered $\beta-$chains (see Appendix)~\cite{Higashinaka03b,Yoshida04a,Fennell05a}. This 2 in - 2 out state is noted ODSI$_{[110]}$.

By ordering the $\alpha-$chains the field hinders thermal fluctuations which improved the thermalization of simulations, even if the presence of modulated phases could not be completely ruled out for all boundaries of the four-dimensional parameter space. The $T=0$ phase diagram of the simulated phases was then computed by Ewald summation (see Appendix and Fig.~\ref{fig:PDh}).

Because of the intrinsic quasi-degeneracy of the 2 in - 2 out configurations, the boundary with the DL phase only barely shifts as the ODSI order quickly gives way to ODSI$_{[110]}$ upon increasing $h$. The AIAO phase, on the other hand, is suppressed by the [110] field and gives way to the DL phase. \textit{The electrically induced double-layer structure is thus strengthened by the [110] magnetic field} to larger $J$ and up to $D_{m}\approx 1.66 \, D_{e}$. For $D_{m}> 1.66 \, D_{e}$, the effective Coulomb repulsion between same-sign monopoles breaks the DL phase in favour of a zincblende or monolayer (ML) structure of monopoles with saturated magnetisation along the [111] direction.\\


\paragraph{Multiferroicity in rare-earth oxides --\hspace{-0.2cm}}our results provide a clear signature of what to look for in experiments and remarkably, this double-layer structure has indeed been previously observed in Tb$_{2}$Ti$_{2}$O$_{7}$ under an external [110] magnetic field~\cite{Ruff10a,Sazonov12a} !

But is Tb$_{2}$Ti$_{2}$O$_{7}$ a good candidate for our theory ? We believe so for three reasons. Firstly, the Ising anisotropy of Tb$^{3+}$ ions~\cite{Gingras00a,Cao09b} and the pinch points observed in polarized neutron scattering~\cite{Fennell12a,Petit12a,Guitteny13a} are strong indications for underlying spin-ice physics. Also, Tb$_{2}$Ti$_{2}$O$_{7}$ possesses a giant magnetostriction~\cite{Aleksandrov85a,Mamsurova86a}, especially along the [110] direction~\cite{Ruff10b}, and the stability of the low temperature spin liquid phase has been recently ascribed to spin-phonon hybridization of the excitations~\cite{Fennell14a}, via dynamical Jahn-Teller coupling~\cite{Bonville14a}. Last but not least, while sample dependence seems to be an issue in this compound, the double-layer structure in a large [110] field has been confirmed by independent experiments~\cite{Ruff10a,Sazonov12a} and is thus a robust feature of Tb$_{2}$Ti$_{2}$O$_{7}$. It should indeed be noted that the nature of the zero- and low-field phases of this material remain under debate -- possibly because of light stuffing/dilution~\cite{Taniguchi13a} -- alternatively described as spin liquid~\cite{Gardner99a,Gardner03a,Bonville11a,Fennell12a,Petit12a} or glassy~\cite{Luo01a,Yasui02a,Fritsch14a} with antiferromagnetic correlations~\cite{Yasui01a,Fritsch14a}. Such behaviour is reminiscent of another rare-earth pyrochlore, Yb$_{2}$Ti$_{2}$O$_{7}$: while being remarkably well parametrized under a high magnetic field~\cite{Ross11a}, the zero field properties of Yb$_{2}$Ti$_{2}$O$_{7}$ noticeably vary between samples~\cite{Yaouanc11a,Ross12a,Ortenzio13a,Chang14a}, also possibly due to light stuffing~\cite{Ross12a}. Given the present low-field uncertainty, our goal here is to propose an alternative scenario for the high field region and to put a new benchmark on the 15-year-old puzzle that is Tb$_{2}$Ti$_{2}$O$_{7}$.

\begin{figure}[ht]
\centering\includegraphics[width=8cm]{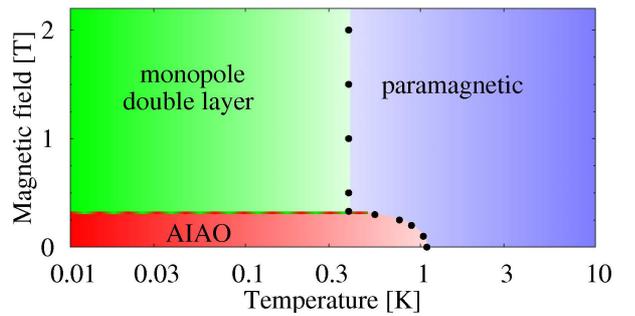}
\caption{Phase diagram parametrized for Tb$_{2}$Ti$_{2}$O$_{7}$ for $J=2.7$ K, $D_{m}=0.48$ K and $D_{e}=0.32$ K. The [110] magnetic field aligns the $\alpha$ chains along the [110] direction, destroying the AIAO order in favour for the double-layer structure. All error bars are smaller than the dots except for the red/green hatched region where simulations were difficult to equilibrate.
}
\label{fig:PDTTO}
\end{figure}

Because of a complex single-ion crystal field~\cite{Bonville11a,Zhang14a,Klekovkina14a,Princep15a}, the effective size of the Tb$^{3+}$ magnetic moments is not fixed, but $\mu = 6\;\mu_{B}$ is a good estimate at low temperature~\cite{Gingras00a} and high field~\cite{Sazonov10a}. With $r_{m}=3.59$\AA ,  Eq.~\ref{eq:NRJscale} gives $D_{m}=0.48$ K. From our theory, the double-layer structure can then be stabilized for $D_{e}^{0}>D_{m}/1.66\approx 0.30$ K, corresponding to an electric moment of $\sim 2. 10^{-31}$ C.m and a displacement of oxygen ions of $\sim 0.6$ pm, which are reasonable estimates for multiferroics~\cite{Khomskii12a,Tokura14a}. Using the parametrization of~\cite{Enjalran04b} with $J=2.7$ K, our multiferroic spin ice model can also explain why the DL phase only appears at finite field~\cite{Sazonov12a,Fritsch14a} (Fig.~\ref{fig:PDTTO}). In light of the sample-dependence issue, it is difficult to push the comparison further to low field, where the phase might be antiferromagnetic but probably not AIAO~\cite{Fritsch14a}, and should be separated from the DL structure by a collective paramagnet up to 2 Tesla. This is where magnetostriction comes into play as a potential complementary facet of our model.

The [110] direction for a magnetic field has been shown to maximize magnetostriction in Tb$_{2}$Ti$_{2}$O$_{7}$, resulting in field-dependent oxygen displacement~\cite{Ruff10b}. Hence, an external magnetic field stabilizes the DL structure not only by suppressing the AIAO order (Fig.~\ref{fig:PDh}), but possibly also by increasing the electric energy scale $D_{e}$. In that case, an even wider range of parameters (\textit{e.g.} $J=0.96$ K~\cite{Mirebeau07a}) and perturbations will lead to a double-layer phase at high field. The possibility to include disorder, anisotropic~\cite{Curnoe07a,Curnoe08a,Bonville14a} or next-nearest-neighbour interactions~\cite{Mcclarty14a} and quantum fluctuations~\cite{Molavian07a,Molavian09a,Mcclarty14a} to our multiferroic spin ice model opens a rich diversity of potential phases to account for the yet uncertain phase of Tb$_{2}$Ti$_{2}$O$_{7}$ in low field. In particular it is tempting to speculate whether the glassy behaviour observed in some samples~\cite{Luo01a,Yasui02a,Fritsch14a} might be a consequence of the dynamically difficult nucleation process discussed in this paper, especially if Tb$_{2}$Ti$_{2}$O$_{7}$ lies close to a spin liquid phase~\cite{Gardner99a,Gardner03a,Bonville11a,Fennell12a,Petit12a,Mirebeau02a}.\\

\paragraph{Conclusion --\hspace{-0.2cm}} in summary, we have shown how magneto-electric coupling can lift the degeneracy of a spin liquid by creating interactions between topological excitations. In spin ice these excitations condense into a bi-layered monopole crystal strengthened by a [110] magnetic field, a non-trivial example of ``magnetic crystallography''. Our theory offers a simple and robust explanation for the ordering of Tb$_{2}$Ti$_{2}$O$_{7}$ in a large [110] field for a reasonably small electric energy scale $D_{e}$.

If we look at the diverse physics emerging from itinerant electrons coupled to spin ice (anomalous and spontaneous Hall effects in Nd$_{2}$Mo$_{2}$O$_{7}$~\cite{Taguchi01a} and Pr$_{2}$Ir$_{2}$O$_{7}$~\cite{Machida10a},  non-Kondo resistivity minimum~\cite{Nakatsuji06a,Sakata11a,Udagawa12a,Chern13a} and a new kind of quantum criticality~\cite{Savary14a} in iridates), we should expect the coupling to an additional, ferroic, degree of freedom to bring a new flavor to spin ice and spin liquids, both at equilibrium and dynamically~\cite{Sarkar14a}. Experiments on Dy$_{2}$Ti$_{2}$O$_{7}$ and Ho$_{2}$Ti$_{2}$O$_{7}$ already suggest the presence of magneto-electric effects~\cite{Katsufuji04a,Saito05a,Liu13a,Grams14a}, which could be enhanced or even qualitatively modified by doping and chemical pressure~\cite{Zhou12a,Lin13a,Xu14a} or with the inclusion of an electric field.
More generally, multiferroicity offers a promising mechanism to control topological defects in magnets. We hope our work will motivate further theoretical and experimental investigations of multiferroic effects in pyrochlores and spin liquids.\\

\begin{acknowledgments}
The authors are thankful to Pascal Qu\'emerais for collaborations at an early stage of this project, and to Owen Benton, Bruce Gaulin, Isabelle Mirebeau, Yukitoshi Motome, Karlo Penc and Oleg Tchernyshyov for useful discussions. This work was supported by funding from the Theory of Quantum Matter unit of the Okinawa Institute of Science and Technology Graduate University.
\end{acknowledgments}


\section{Appendix}

\subsection{Ground states energies}

The $T=0$ phase diagram of Figs.~1 and~4 in the paper was calculated based on the following energies per number of spins $N$. The spin configurations are given in figures~\ref{fig:DML},~\ref{fig:ODSI} and~\ref{fig:AIAO}. 

\begin{align}
\label{eq:NRJ1}
\textrm{AIAO:}\;&E\=+4.09\;D_{m}& &-J\\
\label{eq:NRJ2}
\textrm{DL:}\;&E\=0.0455\;D_{m}\!\!\!\!&-h/\sqrt{6}&\-0.692\;D_{e}\\
\label{eq:NRJ3}
\textrm{ODSI:}\;&E\=-1.95\;D_{m}& &\+J/3\\
\label{eq:NRJ4}
\textrm{ODSI}_{[110]}{\rm :}\;&E\=-1.90\;D_{m}\!\!\!\!&-h\sqrt{6}&\+J/3\\
\textrm{ML:}\;&E\=-0.370\;D_{m}\!\!\!\!&-h\sqrt{6}
\label{eq:NRJ5}
\end{align}

One sees immediately that the DL phase wins over the ML one for $D_{e}>0.60 D_{m}$, and that it is stable for $J/D_{e}\in[-2.07;0.692]$ when $D_{m}=h=0$. While the prefactors for the coupling $J$ and the Zeeman term are straightforward to calculate, the remaining terms require Ewald summation for a precise estimate. However, many of them can actually be calculated analytically to a good approximation~\cite{Castelnovo08a,Brooks14a}. Since it provides a useful insight into the analogy between dipoles and monopoles, we briefly explain the method in the next section. Please note these results are obtained in absence of demagnetization factor. We recover the same values as Yoshida \textit{et al.}~\cite{Yoshida04a} for the magnetic energies of ODSI and ODSI$_{[110]}$ when including the demagnetization factor of a sphere in vacuum.

\begin{figure*}
\centering\includegraphics[width=7cm]{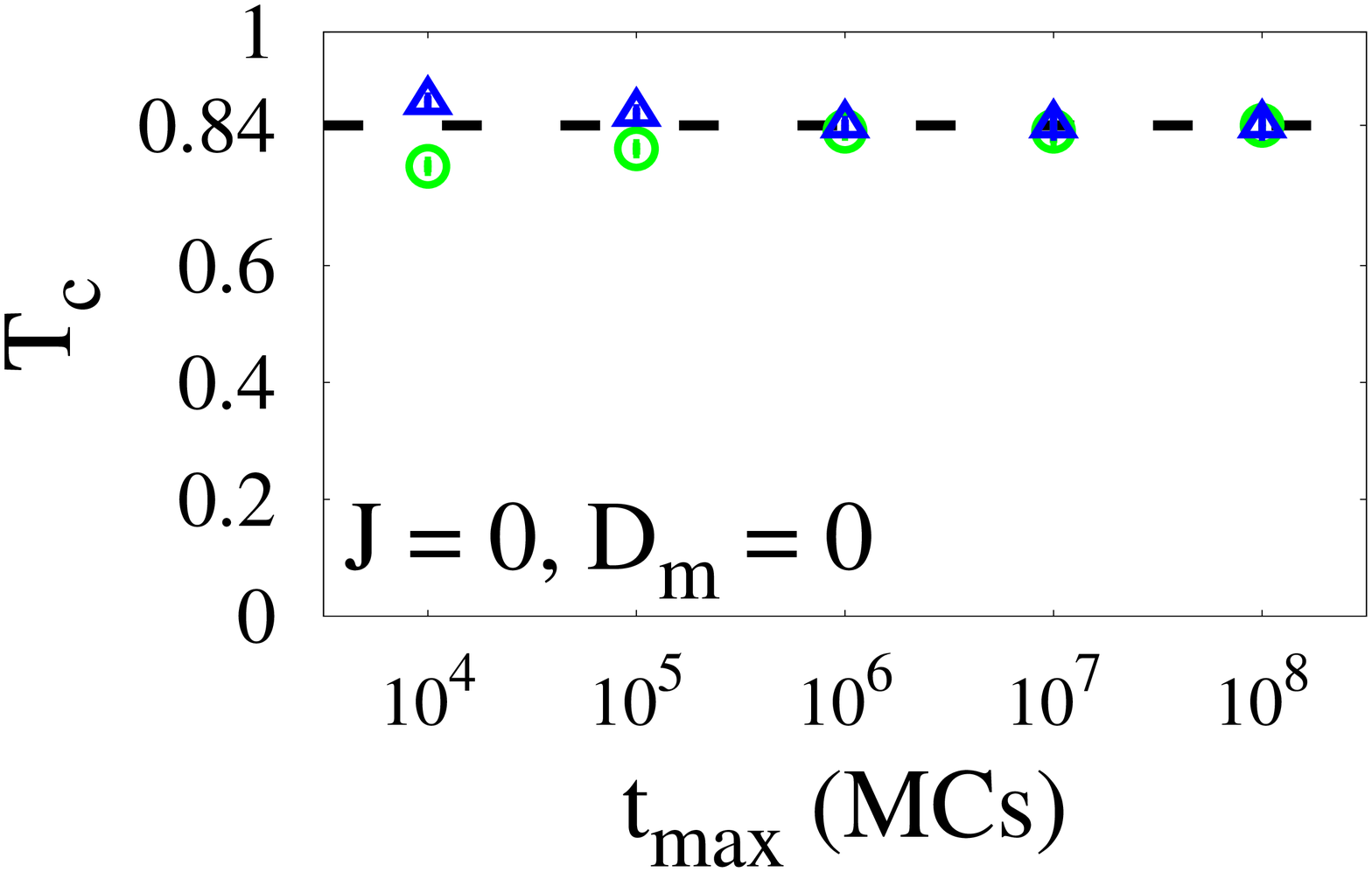}\hspace{1cm}
\centering\includegraphics[width=7cm]{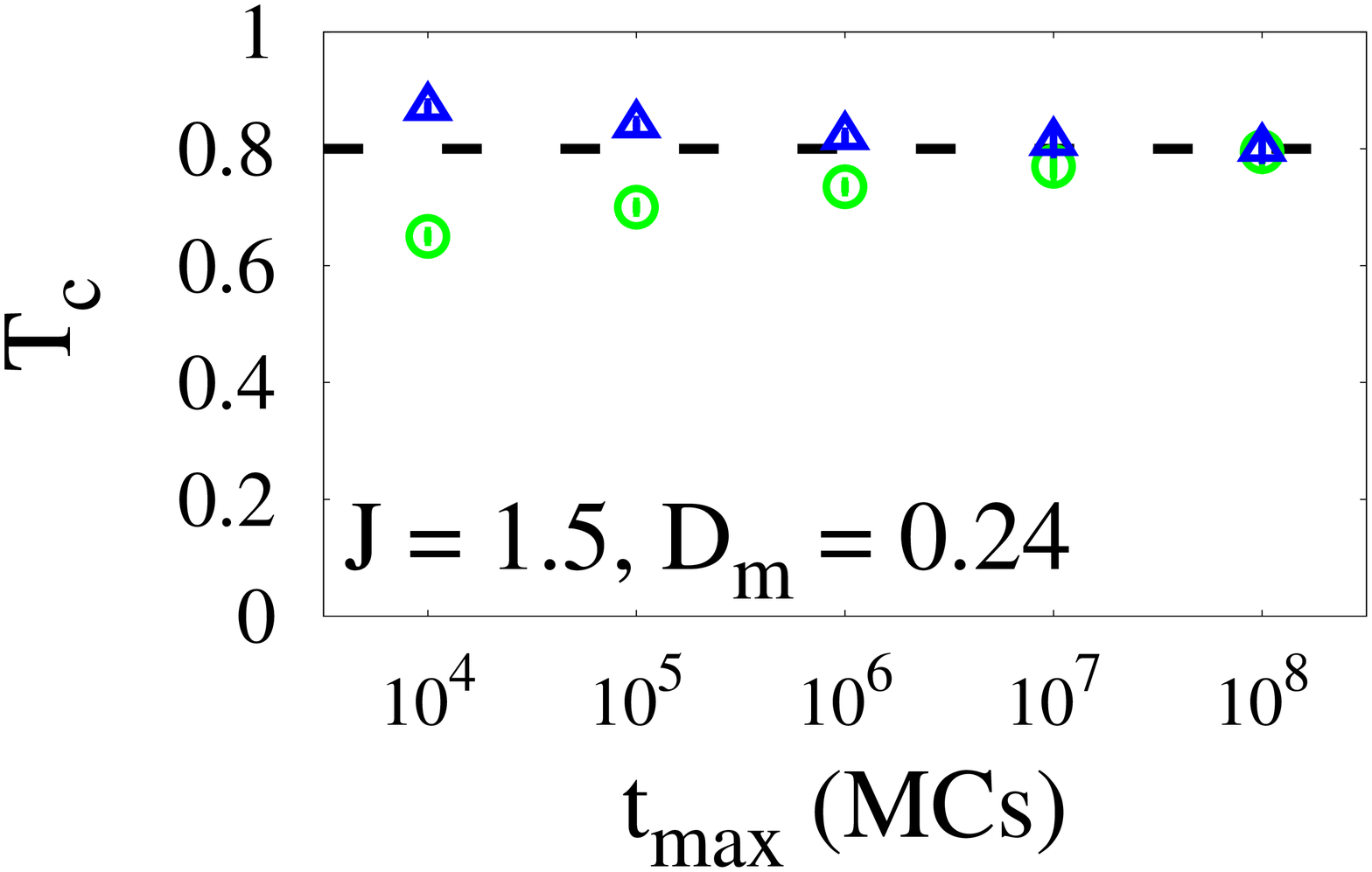}\\
\centering\includegraphics[width=7cm]{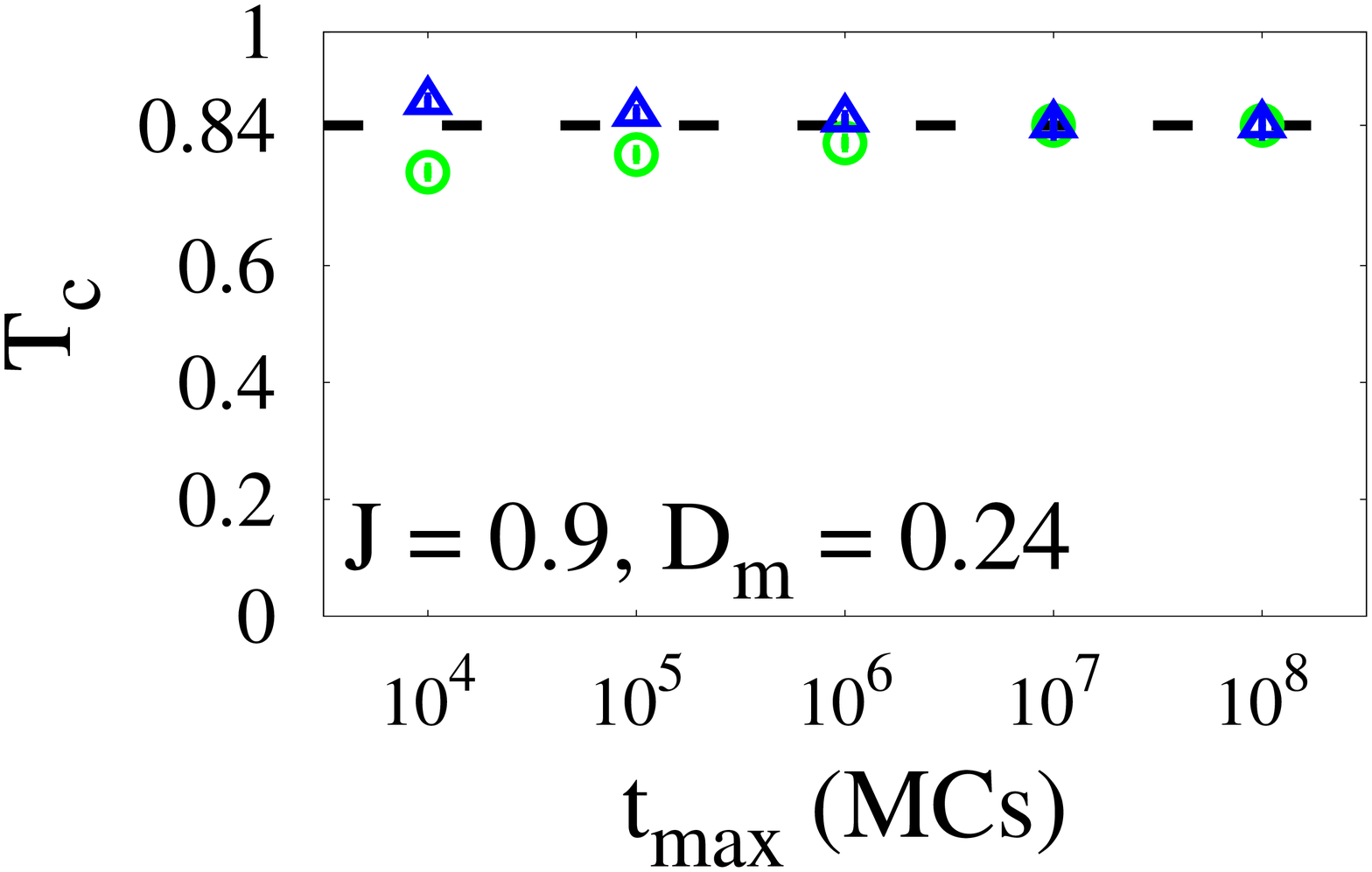}\hspace{1cm}
\centering\includegraphics[width=7cm]{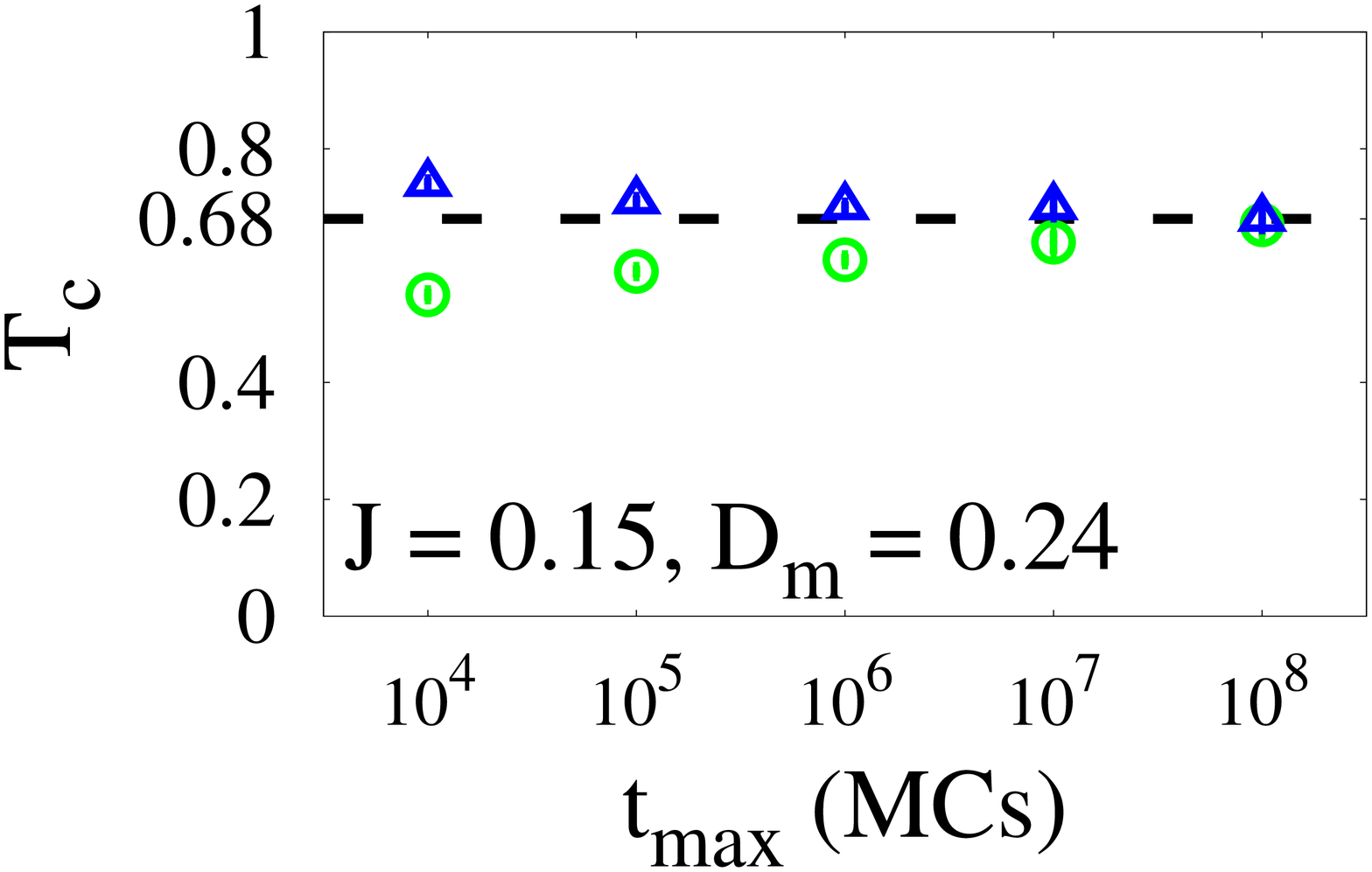}
\caption{
Convergence of the transition temperature as a function of measurement time $t_{\rm max}$ for both equilibration processes (i -- green) and (ii -- blue). For all figures $D_{e}=1$, $h=0$ and $N=432$. We confirm that even if the process is very slow, simulations converge to the same transition temperature indicated by the dashed line.
}
\label{fig:FSStX}
\end{figure*}
\begin{figure*}
\centering\includegraphics[width=7cm]{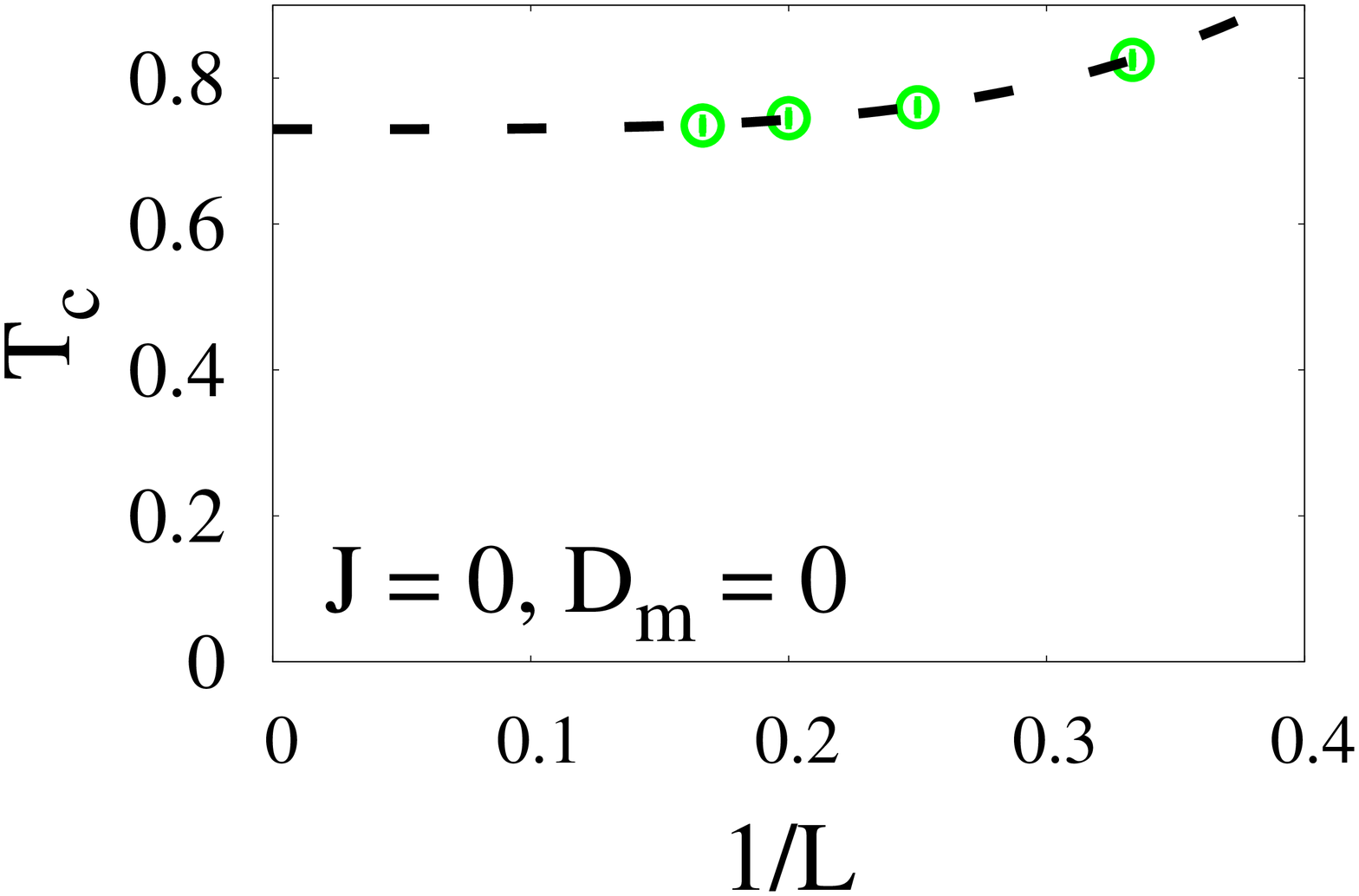}\hspace{1cm}
\centering\includegraphics[width=7cm]{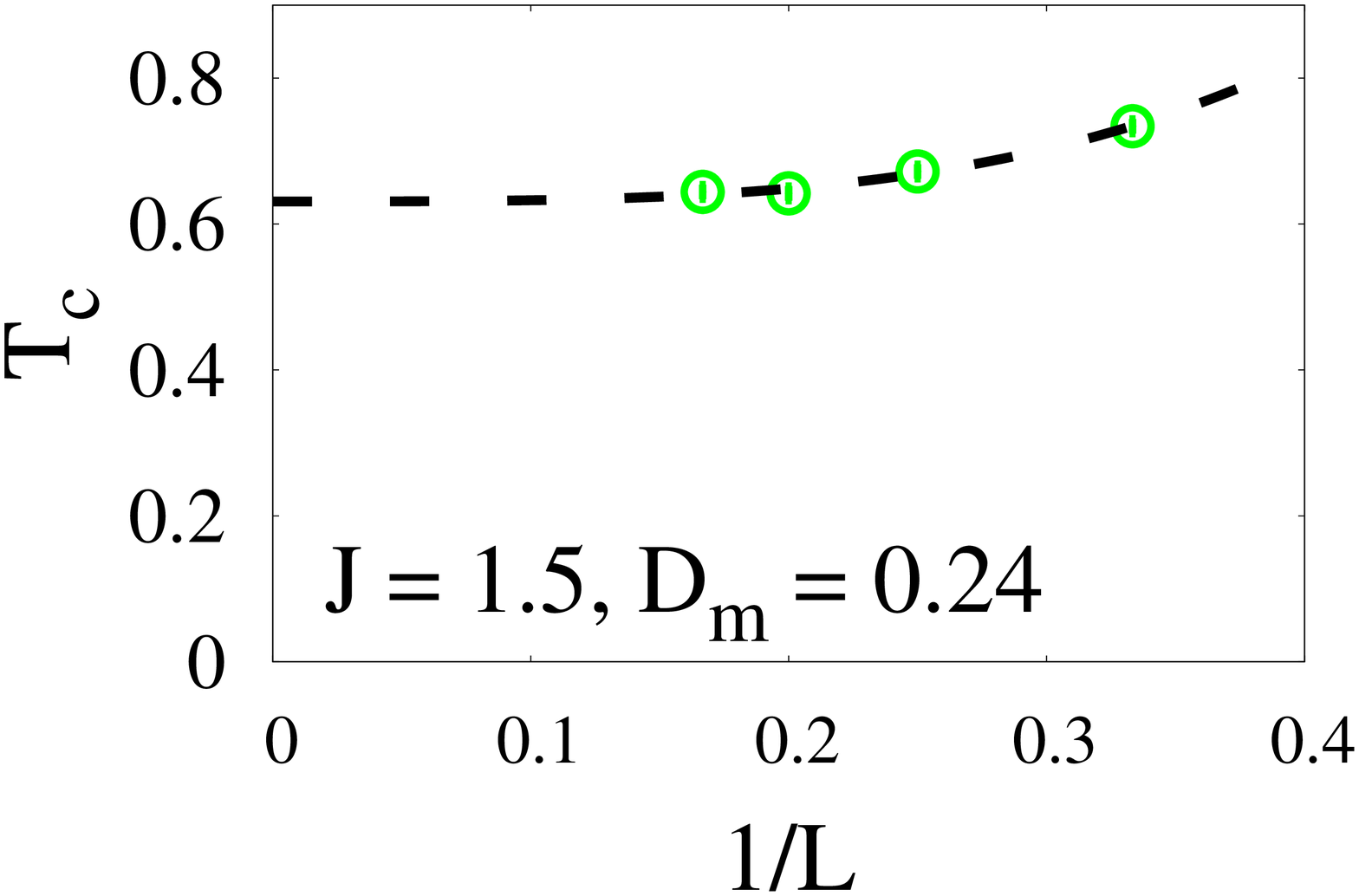}\\
\centering\includegraphics[width=7cm]{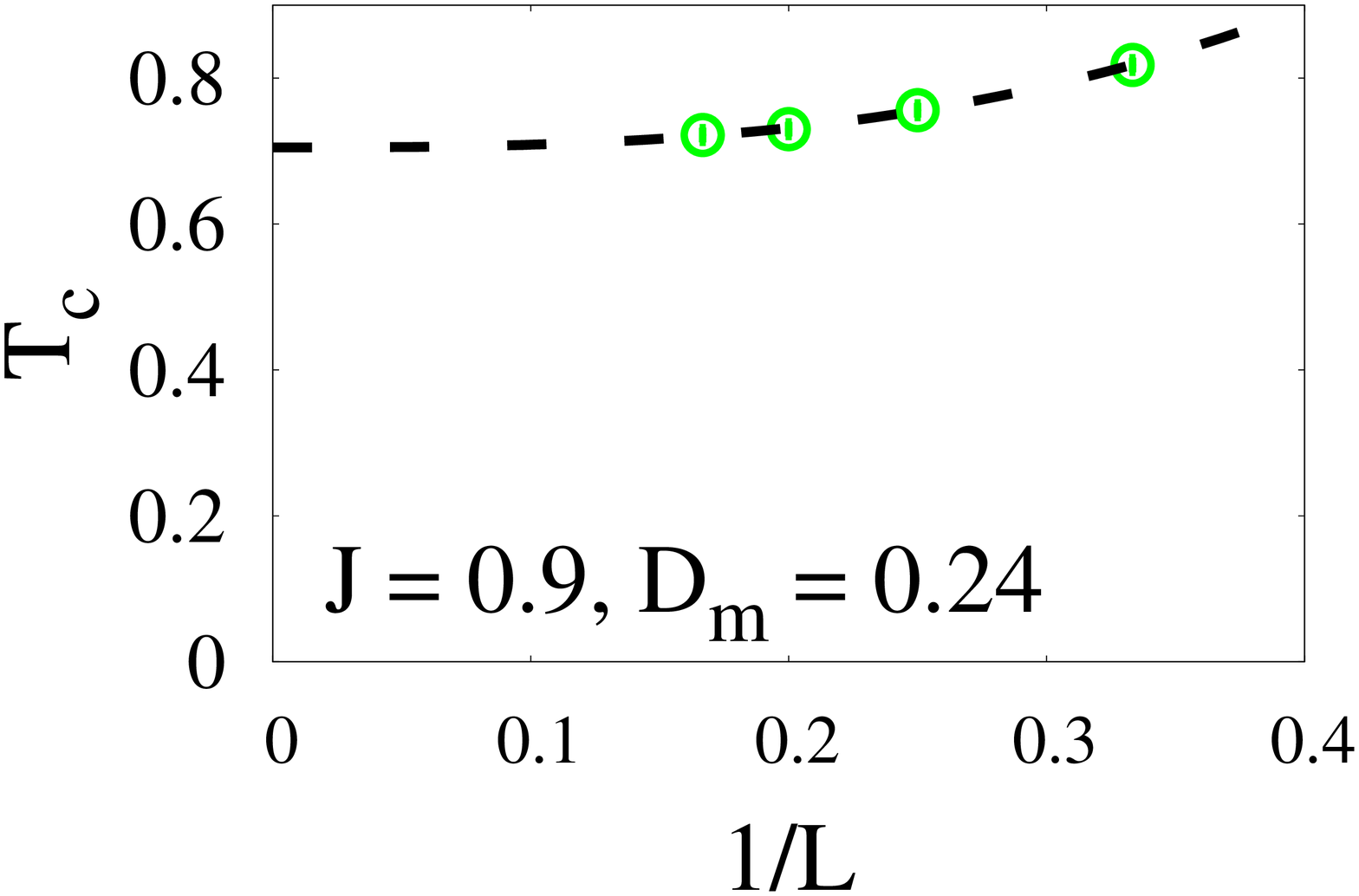}\hspace{1cm}
\centering\includegraphics[width=7cm]{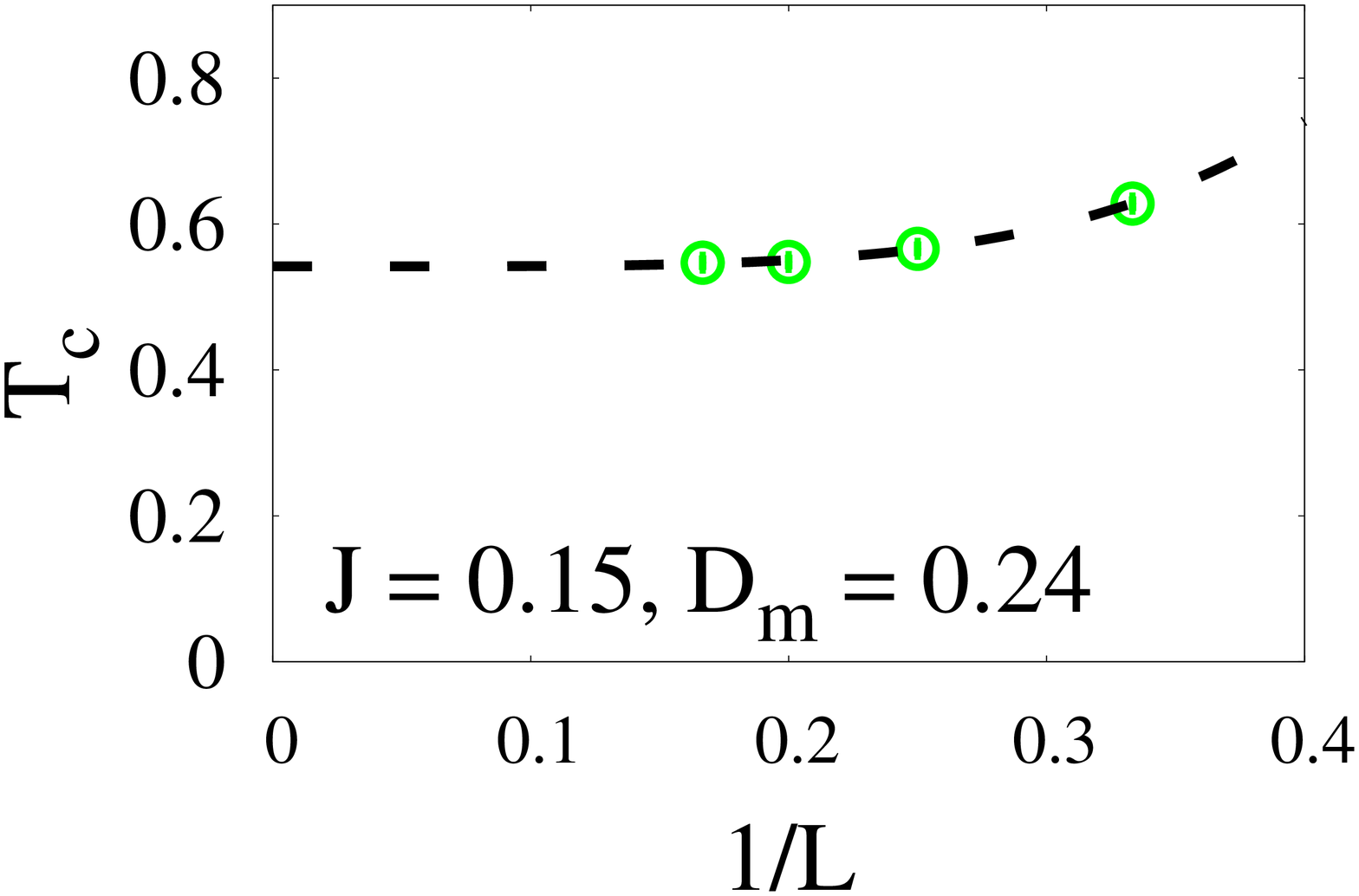}
\caption{Finite size scaling of the transition temperature for equilibration process (i). For all figures $D_{e}=1$, $h=0$ and $L=\{3,4,5,6\}$. The number of pyrochlore sites is $N=16 L^{3}$. The dotted line is the best fit of the form $a+b/L^{c}$, where $a,b,c$ are fitting parameters.}
\label{fig:FSSLX}
\end{figure*}

\subsection{Analytical calculation of the energies}
Let $Q=n\,q_{m}$ denote the magnetic charge on a given diamond site where $q_{m}=2\mu/r_{e}$ and $n\in\{-2,-1,0,1,2\}$. In presence of magnetic dipolar interactions only, the energy cost $p_{n}$ to create a monopole of charge $n$ is (see Supplementary Informations of~\cite{Castelnovo08a})
\begin{eqnarray}
p_{n}=-\frac{8}{3}\left(1+\sqrt{\frac{2}{3}}\right)n^{2}\;D_{m}
\label{eq:chemicalpotential}
\end{eqnarray}

The magnetic interaction between charges is difficult to calculate for any random configuration, but if the system is charge ordered then it is possible to use the Madelung constant of the corresponding crystal structure in order to calculate its Coulomb energy~\cite{Brooks14a}. Both AIAO and ML configurations are ordered in the zincblende structure with Madelung constant $\alpha_{zb}=1.638$. With $N/2$ diamond sites, the Coulomb energy is
\begin{eqnarray}
U_{n}^{c}=-\frac{1}{2}\frac{N}{2}\alpha_{zb}n^{2}\frac{\mu_{0}q_{m}^{2}}{4\pi r_{e}}=-N\alpha_{zb}\frac{2}{3}\sqrt{\frac{2}{3}}n^{2}D_{m}.
\label{eq:Mdlg}
\end{eqnarray}
giving a total energy $E_{n}=U_{n}^{c}-p_{n}\frac{N}{2}$
\begin{eqnarray}
E_{n}&=&\frac{2N}{3}\left(2+(2-\alpha_{zb})\sqrt{\frac{2}{3}}\right)n^{2}\;D_{m}\\
&=&1.53\,n^{2}\,D_{m}
\label{eq:AIAOMLNRJ}
\end{eqnarray}
where $n=1$ for the ML and $n=2$ for AIAO. This value should be compared with a vacuum of charges, \ie the Coulomb spin liquid. Since the degeneracy of the Coulomb spin liquid is weakly lifted, the choice for a reference energy is somewhat arbitrary. With a lowest energy state at $-1.95 D_{m}$ (ODSI) and the highest energy one at $-1.85 D_{m}$ (fully saturated in the [001] direction), we choose $E_{\rm ref}=-1.90D_{m}$ as reference energy and obtain
\begin{eqnarray}
\Delta E_{1}=E_{ML}-E_{\rm ref}=1.53 D_{m} &{\rm vs}& E_{1}=1.53 D_{m}\nonumber\\
\Delta E_{2}=E_{AIAO}-E_{\rm ref}=5.99 D_{m} &{\rm vs}& E_{2}=6.12 D_{m}\nonumber
\label{eq:compare}
\end{eqnarray}
which are in remarkably good agreement. As for the double layer structure, we could not find the Madelung constant for such charge ordering. So we calculated it using Ewald summation for Coulomb interactions and obtained $\alpha_{DL}=0.976$. This gives $E_{DL, \rm Madelung}= 1.89 D_{m}$, to be compared with $\Delta E_{DL}=E_{DL}-E_{\rm ref}=1.95 D_{m}$, within 3\% of error.

\subsection{Finite temperature simulations}

Now that the zero temperature boundary can be determined exactly from equations~(\ref{eq:NRJ1}) to~(\ref{eq:NRJ5}), let us turn our attention to the finite temperature phase diagram, and in particular to the double-layer phase. All simulations were done with parallel tempering, usually with 1 mK difference between parallel temperatures. To further help thermalization, we have developed a variant of the worm algorithm for dipolar spin ice~\cite{Melko04a}, adapted for both electric dipolar interactions and the presence of singly charged monopoles. Measurements were made for a given number of Monte Carlo steps (MCs) noted $t_{max}$.

We used two different equilibration processes to compute the error bars:
\begin{itemize}
\item (i) the system is slowly cooled down from high temperature to the temperature $T$ of measurement during $t_{max}/10$; then it is thermalized at temperature $T$ during $t_{max}/10$; for our model, this method provides a lower bound for the transition temperature.
\item (ii) for any given set of parameters, the ensemble of ground states is known exactly (cf. the previous sections); the system is then quenched into one of the ground state configurations and thermalized at temperature $T$ during $t_{max}/10$; this method is biased towards ordering and offers an upper bound for the transition temperature.
\end{itemize}
These two values provide the error bars plotted on Figs. 1 and 5 of the main text for a system of size $N=1024$ spins. When not visible, the error bars are smaller than the symbols. For the 3-dimensional plot of Fig. 3 of the main text (no error bars), we only used the equilibration process (i) for a system of size $N=432$, which provides a lower estimate of the stability of the DL phase. For Figs. 1, 3 and 5 of the main text, we used $t_{max}=10^{6}$ MCS, except in the double layer phase of Fig. 1 where $t_{max}=10^{7}$ MCS.\\

In Fig.~\ref{fig:FSStX}, we show how these two processes converge to the same value for four different sets of parameters ordering in the double-layer phase (the most difficult ordering process in our simulations). Because of the double long-range interactions and the very long time of thermalization, big system sizes are difficult to simulate: Fig.~\ref{fig:FSSLX} displays how the transition temperature converges to a finite value with increasing linear system size $L$.


\subsection{Husimi tree}

However close to the boundaries, especially with the Coulomb spin liquid, if the equilibration process (ii) always orders in its ground state configuration, the process (i) might not be able to find the true ground state and will be dominated by the neighbouring phase. This is what happens in the hatched regions of Fig.~1 in the main text. In that case, we cannot rely solely on simulations to determine the finite temperature phase diagram and an analytical approach becomes necessary.

The out-of-equilibrium region between the DL and AIAO is rather narrow, which is why we shall focus on the broader one between the Coulomb spin liquid (CSL) and the double layer, by estimating the free energy of the two phases for $D_{m}=h=0$ (cf. yellow/green hatched region in Fig.~1 of the main text).\\

Let us first consider the double layer phase. According to Eq.~\ref{eq:NRJ2}, its internal energy is $U_{DL}=-0.692 D_{e}$. As for the entropy $S_{DL}$, since the transition is strongly first order, fluctuations can be neglected in a first approximation when compared with the Coulomb spin liquid of extensive degeneracy. Thus the free energy is $F_{DL}=U_{DL}-T\,S_{DL}=-0.692$ if we fix $D_{e}=1$.\\

As for the Coulomb spin liquid, since $D_{m}=h=0$, it corresponds to the canonical nearest neighbour spin ice model with electric interactions only between singly-charged monopoles. To obtain an upper and lower estimate of the boundary between the Coulomb spin liquid and the DL phase, we consider the two following cases
\begin{itemize}
\item 1) electric interactions between the dilute monopoles are modeled by an effective chemical potential: the biggest interaction energy gained by a pair of monopoles being $-2D_{e}/3$, we estimate an effective chemical potential of $-D_{e}/3$ per monopole.
\item 2) electric interactions are neglected: since they tend to lower the interacting energies of the dilute monopoles, this should give an upper estimate of the transition temperature.
\end{itemize}
The free energies of both cases can be calculated in the Husimi tree approximation
\begin{eqnarray}
\label{eq:FCSL1}
F_{CSL\,1}&=&-\frac{T}{2}\log\left[\frac{4\,{\rm e}^{\beta D_{e}/3}+{\rm e}^{2\beta J}+3\,{\rm e}^{-2\beta J/3}}{2}\right]\\
F_{CSL\,2}&=&-\frac{T}{2}\log\left[\frac{4+{\rm e}^{2\beta J}+3\,{\rm e}^{-2\beta J/3}}{2}\right]
\label{eq:FCSL2}
\end{eqnarray}
Both free energies reproduce the asymptotic limits of the Coulomb spin liquid entropy, namely the Pauling ($T\rightarrow 0^{+}$) and paramagnetic ($T\rightarrow +\infty$) entropies. When compared with $F_{DL}$, equations~(\ref{eq:FCSL1}) and~(\ref{eq:FCSL2}) provide respectively the lower and upper solid line of Fig.~1 of the main text for $J\in[-2.08:-1.62]$. The estimated extent of the boundary shift due to electric interactions is consistent, within error bars, with our simulations.


\begin{figure}[h]
\centering\includegraphics[width=6cm]{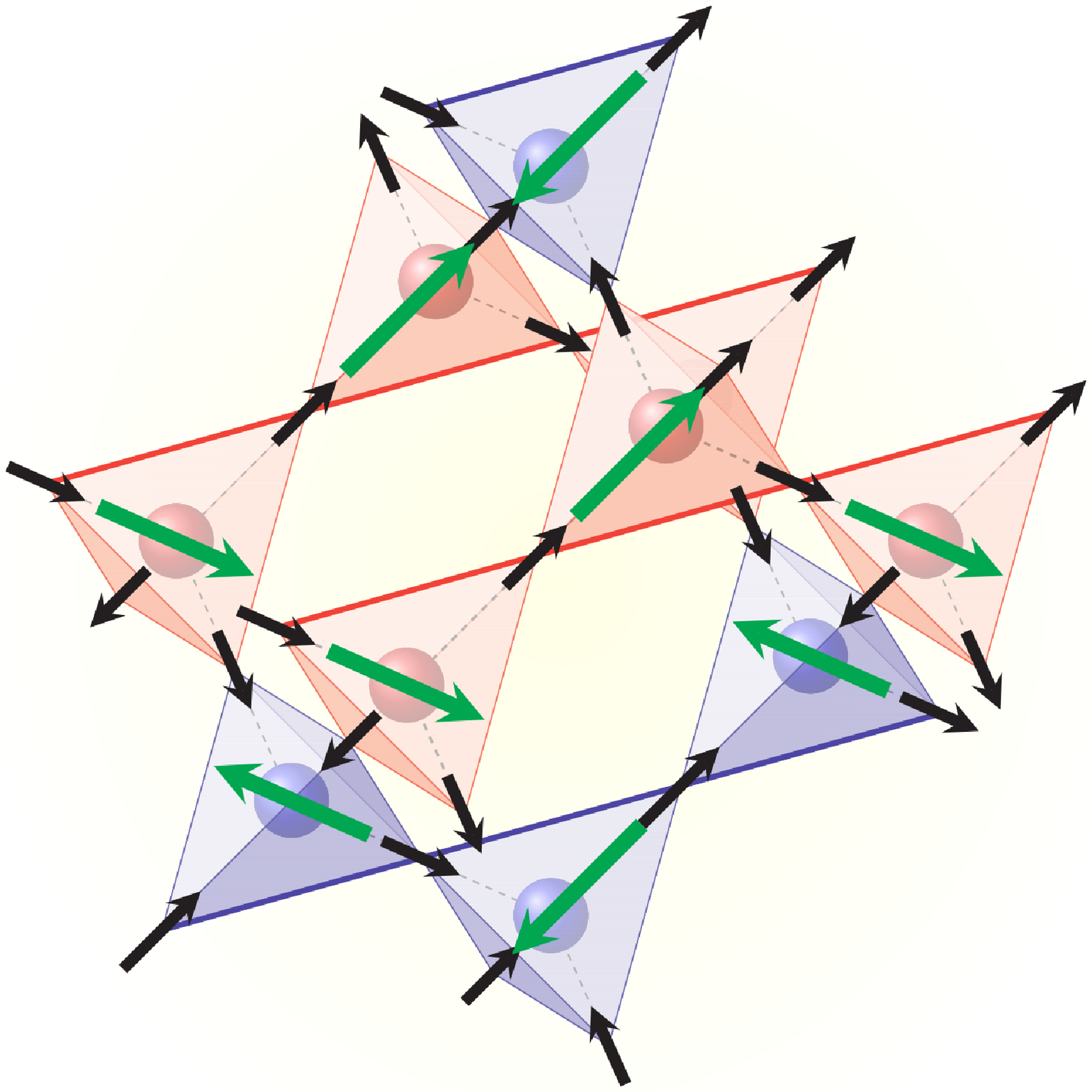}
\centering\includegraphics[width=6cm]{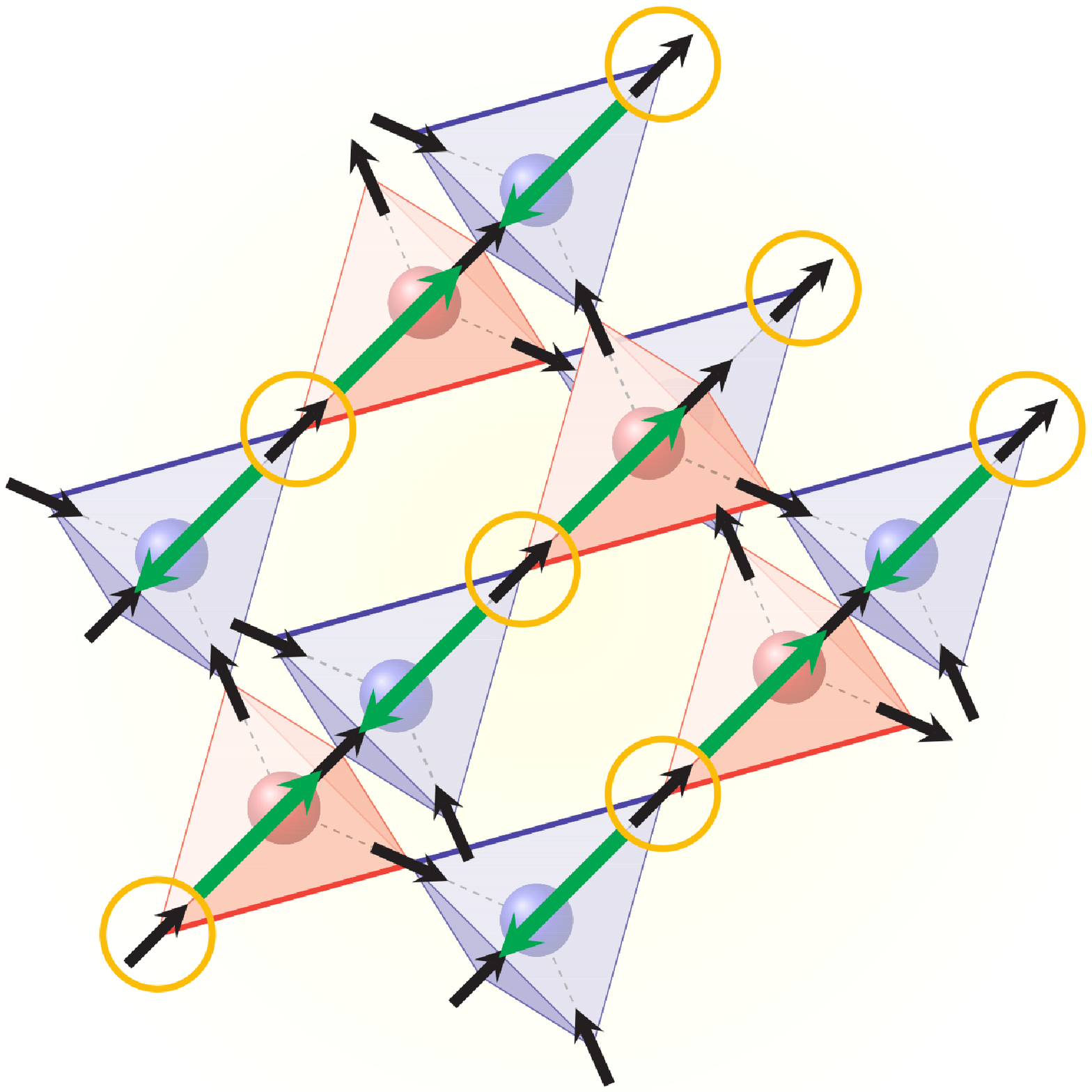}
\caption{
\textbf{Monopole Double Layer} (DL) (\textit{top}) and \textbf{Monopole Mono Layer in a [110] field} (ML) (\textit{bottom}). All tetrahedra carry a single magnetic charge. The green arrows are the electric polarization carried by each monopole. The thick lines are the $\alpha-$chains of spins along the [110] direction. There is no global polarization but there is a saturated magnetization along the $\alpha-$chains in both configurations. For a 3 in - 1 out configuration (resp. 3 out - 1 in), the minority spin is the outward (resp. inward) spin. In the ML phase, each minority spin (circled in orange) is a minority spin for both adjacent monopoles.
}
\label{fig:DML}
\end{figure}
\begin{figure}[ht]
\centering\includegraphics[width=6cm]{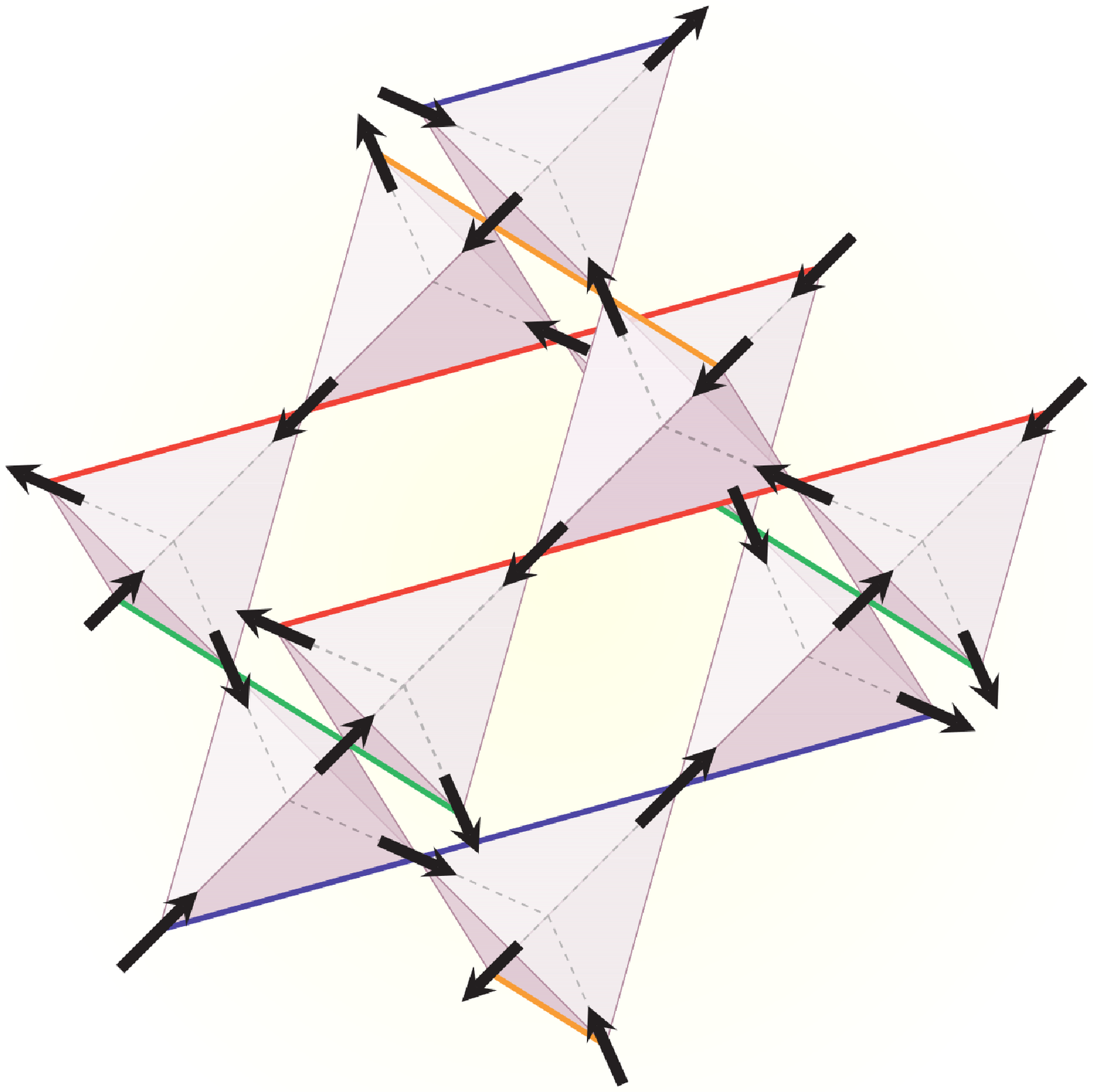}
\centering\includegraphics[width=6cm]{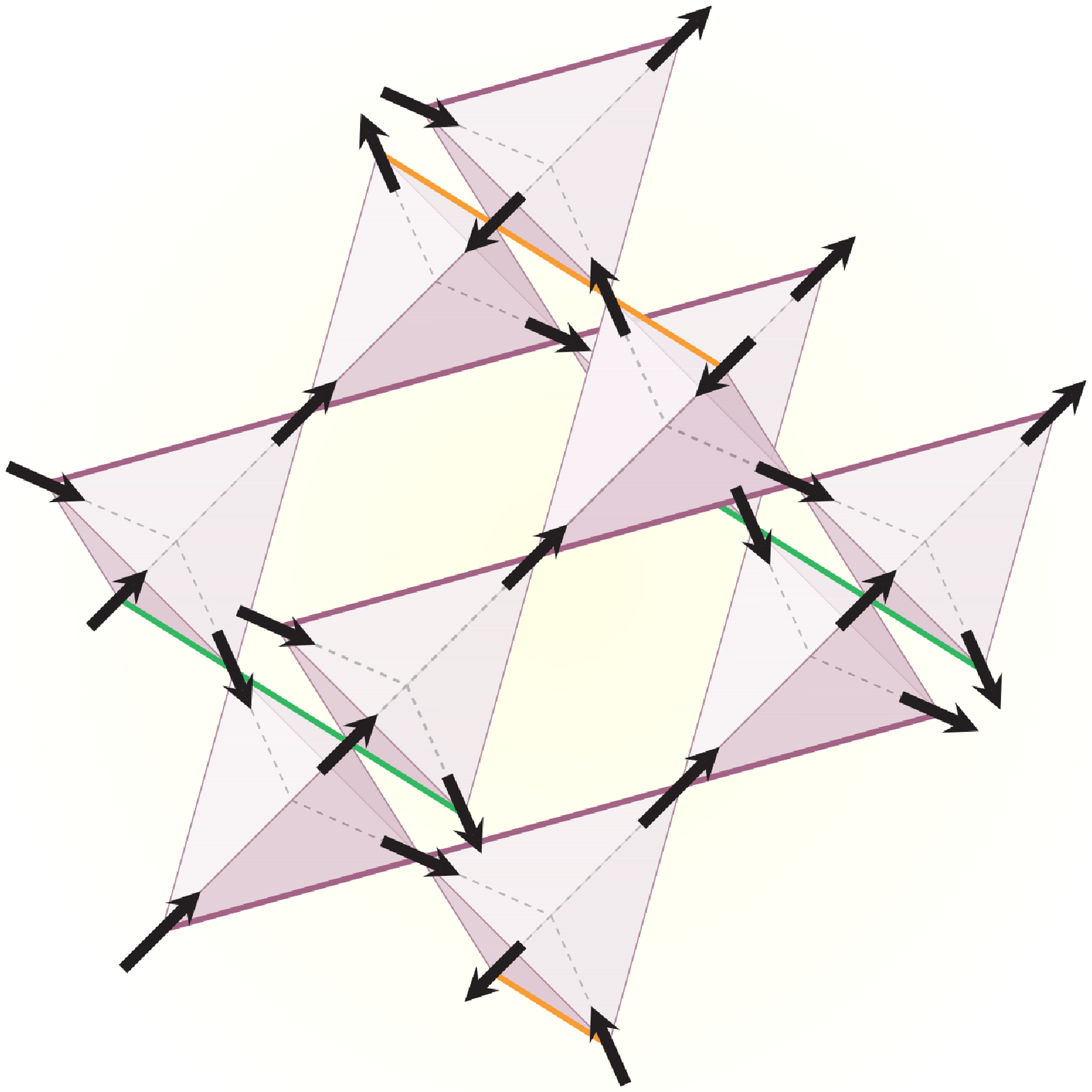}
\caption{
\textbf{Ordered Dipolar Spin Ice} (ODSI) (\textit{top}) and \textbf{Ordered Dipolar Spin Ice in a [110] field} (ODSI$_{[110]}$) (\textit{bottom}). All tetrahedra are in the 2 in - 2 out configuration, \textit{i.e.} without electric polarization. For the ODSI phase, all spins within a (001) plane point in the same direction. The magnetisation from one (001) plane to the next rotates by $\pi/2$. There is no global magnetisation. The ODSI$_{[110]}$ configurations have the same structure as for ODSI except that all spins on the $\alpha-$ chains (thick violet bonds) point in the same direction, giving rise to a saturated magnetization. 
}
\label{fig:ODSI}
\end{figure}
\begin{figure}[h]
\centering\includegraphics[width=6cm]{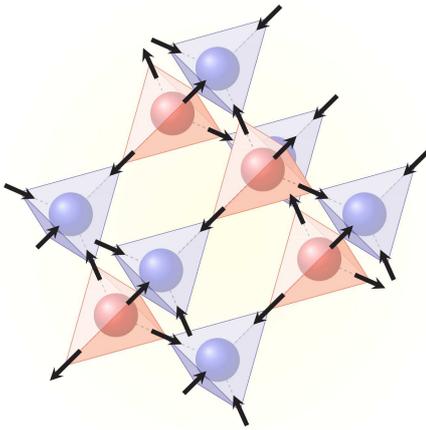}
\caption{\textbf{All in / All out }(AIAO). There is no electric polarization and no global magnetization.}
\label{fig:AIAO}
\end{figure}

\bibliographystyle{apsrev4-1}
\bibliography{../../../../biblio}

\begin{thebibliography}{84}%
\makeatletter
\providecommand \@ifxundefined [1]{%
 \@ifx{#1\undefined}
}%
\providecommand \@ifnum [1]{%
 \ifnum #1\expandafter \@firstoftwo
 \else \expandafter \@secondoftwo
 \fi
}%
\providecommand \@ifx [1]{%
 \ifx #1\expandafter \@firstoftwo
 \else \expandafter \@secondoftwo
 \fi
}%
\providecommand \natexlab [1]{#1}%
\providecommand \enquote  [1]{``#1''}%
\providecommand \bibnamefont  [1]{#1}%
\providecommand \bibfnamefont [1]{#1}%
\providecommand \citenamefont [1]{#1}%
\providecommand \href@noop [0]{\@secondoftwo}%
\providecommand \href [0]{\begingroup \@sanitize@url \@href}%
\providecommand \@href[1]{\@@startlink{#1}\@@href}%
\providecommand \@@href[1]{\endgroup#1\@@endlink}%
\providecommand \@sanitize@url [0]{\catcode `\\12\catcode `\$12\catcode
  `\&12\catcode `\#12\catcode `\^12\catcode `\_12\catcode `\%12\relax}%
\providecommand \@@startlink[1]{}%
\providecommand \@@endlink[0]{}%
\providecommand \url  [0]{\begingroup\@sanitize@url \@url }%
\providecommand \@url [1]{\endgroup\@href {#1}{\urlprefix }}%
\providecommand \urlprefix  [0]{URL }%
\providecommand \Eprint [0]{\href }%
\providecommand \doibase [0]{http://dx.doi.org/}%
\providecommand \selectlanguage [0]{\@gobble}%
\providecommand \bibinfo  [0]{\@secondoftwo}%
\providecommand \bibfield  [0]{\@secondoftwo}%
\providecommand \translation [1]{[#1]}%
\providecommand \BibitemOpen [0]{}%
\providecommand \bibitemStop [0]{}%
\providecommand \bibitemNoStop [0]{.\EOS\space}%
\providecommand \EOS [0]{\spacefactor3000\relax}%
\providecommand \BibitemShut  [1]{\csname bibitem#1\endcsname}%
\let\auto@bib@innerbib\@empty
\bibitem [{\citenamefont {Kimura}\ \emph {et~al.}(2003)\citenamefont {Kimura},
  \citenamefont {Goto}, \citenamefont {Shintani}, \citenamefont {Ishizaka},
  \citenamefont {Arima},\ and\ \citenamefont {Tokura}}]{Kimura03a}%
  \BibitemOpen
  \bibfield  {author} {\bibinfo {author} {\bibfnamefont {T.}~\bibnamefont
  {Kimura}}, \bibinfo {author} {\bibfnamefont {T.}~\bibnamefont {Goto}},
  \bibinfo {author} {\bibfnamefont {H.}~\bibnamefont {Shintani}}, \bibinfo
  {author} {\bibfnamefont {K.}~\bibnamefont {Ishizaka}}, \bibinfo {author}
  {\bibfnamefont {T.}~\bibnamefont {Arima}}, \ and\ \bibinfo {author}
  {\bibfnamefont {Y.}~\bibnamefont {Tokura}},\ }\href {\doibase
  10.1038/nature02018} {\bibfield  {journal} {\bibinfo  {journal} {Nature}\
  }\textbf {\bibinfo {volume} {426}},\ \bibinfo {pages} {55} (\bibinfo {year}
  {2003})}\BibitemShut {NoStop}%
\bibitem [{\citenamefont {Hur}\ \emph {et~al.}(2004)\citenamefont {Hur},
  \citenamefont {Park}, \citenamefont {Sharma}, \citenamefont {Ahn},
  \citenamefont {Guha},\ and\ \citenamefont {Cheong}}]{Hur04a}%
  \BibitemOpen
  \bibfield  {author} {\bibinfo {author} {\bibfnamefont {N.}~\bibnamefont
  {Hur}}, \bibinfo {author} {\bibfnamefont {S.}~\bibnamefont {Park}}, \bibinfo
  {author} {\bibfnamefont {P.}~\bibnamefont {Sharma}}, \bibinfo {author}
  {\bibfnamefont {J.}~\bibnamefont {Ahn}}, \bibinfo {author} {\bibfnamefont
  {S.}~\bibnamefont {Guha}}, \ and\ \bibinfo {author} {\bibfnamefont
  {S.}~\bibnamefont {Cheong}},\ }\href {\doibase 10.1038/nature02572}
  {\bibfield  {journal} {\bibinfo  {journal} {Nature}\ }\textbf {\bibinfo
  {volume} {429}},\ \bibinfo {pages} {392} (\bibinfo {year}
  {2004})}\BibitemShut {NoStop}%
\bibitem [{\citenamefont {Katsura}\ \emph {et~al.}(2005)\citenamefont
  {Katsura}, \citenamefont {Nagaosa},\ and\ \citenamefont
  {Balatsky}}]{Katsura05a}%
  \BibitemOpen
  \bibfield  {author} {\bibinfo {author} {\bibfnamefont {H.}~\bibnamefont
  {Katsura}}, \bibinfo {author} {\bibfnamefont {N.}~\bibnamefont {Nagaosa}}, \
  and\ \bibinfo {author} {\bibfnamefont {A.~V.}\ \bibnamefont {Balatsky}},\
  }\href {\doibase 10.1103/PhysRevLett.95.057205} {\bibfield  {journal}
  {\bibinfo  {journal} {Phys. Rev. Lett.}\ }\textbf {\bibinfo {volume} {95}},\
  \bibinfo {pages} {057205} (\bibinfo {year} {2005})}\BibitemShut {NoStop}%
\bibitem [{\citenamefont {Sergienko}\ and\ \citenamefont
  {Dagotto}(2006)}]{Sergienko06a}%
  \BibitemOpen
  \bibfield  {author} {\bibinfo {author} {\bibfnamefont {I.~A.}\ \bibnamefont
  {Sergienko}}\ and\ \bibinfo {author} {\bibfnamefont {E.}~\bibnamefont
  {Dagotto}},\ }\href {\doibase 10.1103/PhysRevB.73.094434} {\bibfield
  {journal} {\bibinfo  {journal} {Phys. Rev. B}\ }\textbf {\bibinfo {volume}
  {73}},\ \bibinfo {pages} {094434} (\bibinfo {year} {2006})}\BibitemShut
  {NoStop}%
\bibitem [{\citenamefont {Mostovoy}(2006)}]{Mostovoy06a}%
  \BibitemOpen
  \bibfield  {author} {\bibinfo {author} {\bibfnamefont {M.}~\bibnamefont
  {Mostovoy}},\ }\href {\doibase 10.1103/PhysRevLett.96.067601} {\bibfield
  {journal} {\bibinfo  {journal} {Phys. Rev. Lett.}\ }\textbf {\bibinfo
  {volume} {96}},\ \bibinfo {pages} {067601} (\bibinfo {year}
  {2006})}\BibitemShut {NoStop}%
\bibitem [{\citenamefont {Khomskii}(2006)}]{Khomskii06a}%
  \BibitemOpen
  \bibfield  {author} {\bibinfo {author} {\bibfnamefont {D.~I.}\ \bibnamefont
  {Khomskii}},\ }\href {\doibase 10.1016/j.jmmm.2006.01.238} {\bibfield
  {journal} {\bibinfo  {journal} {Journal of Magnetism and Magnetic Materials}\
  }\textbf {\bibinfo {volume} {306}},\ \bibinfo {pages} {1} (\bibinfo {year}
  {2006})}\BibitemShut {NoStop}%
\bibitem [{\citenamefont {Cheong}\ and\ \citenamefont
  {Mostovoy}(2007)}]{Cheong07a}%
  \BibitemOpen
  \bibfield  {author} {\bibinfo {author} {\bibfnamefont {S.-W.}\ \bibnamefont
  {Cheong}}\ and\ \bibinfo {author} {\bibfnamefont {M.}~\bibnamefont
  {Mostovoy}},\ }\href {\doibase 10.1038/nmat1804} {\bibfield  {journal}
  {\bibinfo  {journal} {Nature Materials}\ }\textbf {\bibinfo {volume} {6}},\
  \bibinfo {pages} {13} (\bibinfo {year} {2007})}\BibitemShut {NoStop}%
\bibitem [{\citenamefont {Tokura}\ \emph {et~al.}(2014)\citenamefont {Tokura},
  \citenamefont {Seki},\ and\ \citenamefont {Nagaosa}}]{Tokura14a}%
  \BibitemOpen
  \bibfield  {author} {\bibinfo {author} {\bibfnamefont {Y.}~\bibnamefont
  {Tokura}}, \bibinfo {author} {\bibfnamefont {S.}~\bibnamefont {Seki}}, \ and\
  \bibinfo {author} {\bibfnamefont {N.}~\bibnamefont {Nagaosa}},\ }\href
  {http://stacks.iop.org/0034-4885/77/i=7/a=076501} {\bibfield  {journal}
  {\bibinfo  {journal} {Reports on Progress in Physics}\ }\textbf {\bibinfo
  {volume} {77}},\ \bibinfo {pages} {076501} (\bibinfo {year}
  {2014})}\BibitemShut {NoStop}%
\bibitem [{\citenamefont {Gardner}\ \emph {et~al.}(2010)\citenamefont
  {Gardner}, \citenamefont {Gingras},\ and\ \citenamefont
  {Greedan}}]{Gardner10a}%
  \BibitemOpen
  \bibfield  {author} {\bibinfo {author} {\bibfnamefont {J.~S.}\ \bibnamefont
  {Gardner}}, \bibinfo {author} {\bibfnamefont {M.~J.~P.}\ \bibnamefont
  {Gingras}}, \ and\ \bibinfo {author} {\bibfnamefont {J.~E.}\ \bibnamefont
  {Greedan}},\ }\href {\doibase DOI 10.1103/RevModPhys.82.53} {\bibfield
  {journal} {\bibinfo  {journal} {Reviews of Modern Physics}\ }\textbf
  {\bibinfo {volume} {82}},\ \bibinfo {pages} {53} (\bibinfo {year}
  {2010})}\BibitemShut {NoStop}%
\bibitem [{\citenamefont {Shores}\ \emph {et~al.}(2005)\citenamefont {Shores},
  \citenamefont {Nytko}, \citenamefont {Bartlett},\ and\ \citenamefont
  {Nocera}}]{Shores05a}%
  \BibitemOpen
  \bibfield  {author} {\bibinfo {author} {\bibfnamefont {M.~P.}\ \bibnamefont
  {Shores}}, \bibinfo {author} {\bibfnamefont {E.~A.}\ \bibnamefont {Nytko}},
  \bibinfo {author} {\bibfnamefont {B.~M.}\ \bibnamefont {Bartlett}}, \ and\
  \bibinfo {author} {\bibfnamefont {D.~G.}\ \bibnamefont {Nocera}},\ }\href
  {\doibase 10.1021/ja053891p} {\bibfield  {journal} {\bibinfo  {journal}
  {Journal of the American Chemical Society}\ }\textbf {\bibinfo {volume}
  {127}},\ \bibinfo {pages} {13462} (\bibinfo {year} {2005})},\ \Eprint
  {http://arxiv.org/abs/http://dx.doi.org/10.1021/ja053891p}
  {http://dx.doi.org/10.1021/ja053891p} \BibitemShut {NoStop}%
\bibitem [{\citenamefont {Okamoto}\ \emph {et~al.}(2009)\citenamefont
  {Okamoto}, \citenamefont {Yoshida},\ and\ \citenamefont
  {Hiroi}}]{Okamoto09a}%
  \BibitemOpen
  \bibfield  {author} {\bibinfo {author} {\bibfnamefont {Y.}~\bibnamefont
  {Okamoto}}, \bibinfo {author} {\bibfnamefont {H.}~\bibnamefont {Yoshida}}, \
  and\ \bibinfo {author} {\bibfnamefont {Z.}~\bibnamefont {Hiroi}},\ }\href
  {\doibase 10.1143/JPSJ.78.033701} {\bibfield  {journal} {\bibinfo  {journal}
  {Journal of the Physical Society of Japan}\ }\textbf {\bibinfo {volume}
  {78}},\ \bibinfo {pages} {033701} (\bibinfo {year} {2009})},\ \Eprint
  {http://arxiv.org/abs/http://dx.doi.org/10.1143/JPSJ.78.033701}
  {http://dx.doi.org/10.1143/JPSJ.78.033701} \BibitemShut {NoStop}%
\bibitem [{\citenamefont {Quilliam}\ \emph {et~al.}(2011)\citenamefont
  {Quilliam}, \citenamefont {Bert}, \citenamefont {Colman}, \citenamefont
  {Boldrin}, \citenamefont {Wills},\ and\ \citenamefont
  {Mendels}}]{Quilliam11b}%
  \BibitemOpen
  \bibfield  {author} {\bibinfo {author} {\bibfnamefont {J.~A.}\ \bibnamefont
  {Quilliam}}, \bibinfo {author} {\bibfnamefont {F.}~\bibnamefont {Bert}},
  \bibinfo {author} {\bibfnamefont {R.~H.}\ \bibnamefont {Colman}}, \bibinfo
  {author} {\bibfnamefont {D.}~\bibnamefont {Boldrin}}, \bibinfo {author}
  {\bibfnamefont {A.~S.}\ \bibnamefont {Wills}}, \ and\ \bibinfo {author}
  {\bibfnamefont {P.}~\bibnamefont {Mendels}},\ }\href {\doibase
  10.1103/PhysRevB.84.180401} {\bibfield  {journal} {\bibinfo  {journal} {Phys.
  Rev. B}\ }\textbf {\bibinfo {volume} {84}},\ \bibinfo {pages} {180401}
  (\bibinfo {year} {2011})}\BibitemShut {NoStop}%
\bibitem [{\citenamefont {Kanoda}\ and\ \citenamefont
  {Kato}(2011)}]{Kanoda11a}%
  \BibitemOpen
  \bibfield  {author} {\bibinfo {author} {\bibfnamefont {K.}~\bibnamefont
  {Kanoda}}\ and\ \bibinfo {author} {\bibfnamefont {R.}~\bibnamefont {Kato}},\
  }\href {\doibase 10.1146/annurev-conmatphys-062910-140521} {\bibfield
  {journal} {\bibinfo  {journal} {Annual Review of Condensed Matter Physics}\
  }\textbf {\bibinfo {volume} {2}},\ \bibinfo {pages} {167} (\bibinfo {year}
  {2011})}\BibitemShut {NoStop}%
\bibitem [{\citenamefont {Modic}\ \emph {et~al.}(2014)\citenamefont {Modic},
  \citenamefont {Smidt}, \citenamefont {Kimchi}, \citenamefont {Breznay},
  \citenamefont {Biffin}, \citenamefont {Choi}, \citenamefont {Johnson},
  \citenamefont {Coldea}, \citenamefont {Watkins-Curry}, \citenamefont
  {McCandless}, \citenamefont {Chan}, \citenamefont {Gandara}, \citenamefont
  {Islam}, \citenamefont {Vishwanath}, \citenamefont {Shekhter}, \citenamefont
  {McDonald},\ and\ \citenamefont {Analytis}}]{Modic14a}%
  \BibitemOpen
  \bibfield  {author} {\bibinfo {author} {\bibfnamefont {K.~A.}\ \bibnamefont
  {Modic}}, \bibinfo {author} {\bibfnamefont {T.~E.}\ \bibnamefont {Smidt}},
  \bibinfo {author} {\bibfnamefont {I.}~\bibnamefont {Kimchi}}, \bibinfo
  {author} {\bibfnamefont {N.~P.}\ \bibnamefont {Breznay}}, \bibinfo {author}
  {\bibfnamefont {A.}~\bibnamefont {Biffin}}, \bibinfo {author} {\bibfnamefont
  {S.}~\bibnamefont {Choi}}, \bibinfo {author} {\bibfnamefont {R.~D.}\
  \bibnamefont {Johnson}}, \bibinfo {author} {\bibfnamefont {R.}~\bibnamefont
  {Coldea}}, \bibinfo {author} {\bibfnamefont {P.}~\bibnamefont
  {Watkins-Curry}}, \bibinfo {author} {\bibfnamefont {G.~T.}\ \bibnamefont
  {McCandless}}, \bibinfo {author} {\bibfnamefont {J.~Y.}\ \bibnamefont
  {Chan}}, \bibinfo {author} {\bibfnamefont {F.}~\bibnamefont {Gandara}},
  \bibinfo {author} {\bibfnamefont {Z.}~\bibnamefont {Islam}}, \bibinfo
  {author} {\bibfnamefont {A.}~\bibnamefont {Vishwanath}}, \bibinfo {author}
  {\bibfnamefont {A.}~\bibnamefont {Shekhter}}, \bibinfo {author}
  {\bibfnamefont {R.~D.}\ \bibnamefont {McDonald}}, \ and\ \bibinfo {author}
  {\bibfnamefont {J.~G.}\ \bibnamefont {Analytis}},\ }\href {\doibase
  10.1038/ncomms5203} {\bibfield  {journal} {\bibinfo  {journal} {Nature
  Communications}\ }\textbf {\bibinfo {volume} {5}} (\bibinfo {year} {2014}),\
  10.1038/ncomms5203}\BibitemShut {NoStop}%
\bibitem [{\citenamefont {Takayama}\ \emph {et~al.}(2014)\citenamefont
  {Takayama}, \citenamefont {Kato}, \citenamefont {Dinnebier}, \citenamefont
  {Nuss},\ and\ \citenamefont {Takagi}}]{Takayama14a}%
  \BibitemOpen
  \bibfield  {author} {\bibinfo {author} {\bibfnamefont {T.}~\bibnamefont
  {Takayama}}, \bibinfo {author} {\bibfnamefont {A.}~\bibnamefont {Kato}},
  \bibinfo {author} {\bibfnamefont {R.}~\bibnamefont {Dinnebier}}, \bibinfo
  {author} {\bibfnamefont {J.}~\bibnamefont {Nuss}}, \ and\ \bibinfo {author}
  {\bibfnamefont {H.}~\bibnamefont {Takagi}},\ }\href@noop {} {\bibfield
  {journal} {\bibinfo  {journal} {ArXiv e-prints}\ } (\bibinfo {year}
  {2014})},\ \Eprint {http://arxiv.org/abs/1403.3296} {arXiv:1403.3296
  [cond-mat.str-el]} \BibitemShut {NoStop}%
\bibitem [{\citenamefont {Harris}\ \emph {et~al.}(1997)\citenamefont {Harris},
  \citenamefont {Bramwell}, \citenamefont {McMorrow}, \citenamefont {Zeiske},\
  and\ \citenamefont {Godfrey}}]{Harris97a}%
  \BibitemOpen
  \bibfield  {author} {\bibinfo {author} {\bibfnamefont {M.~J.}\ \bibnamefont
  {Harris}}, \bibinfo {author} {\bibfnamefont {S.~T.}\ \bibnamefont
  {Bramwell}}, \bibinfo {author} {\bibfnamefont {D.~F.}\ \bibnamefont
  {McMorrow}}, \bibinfo {author} {\bibfnamefont {T.}~\bibnamefont {Zeiske}}, \
  and\ \bibinfo {author} {\bibfnamefont {K.~W.}\ \bibnamefont {Godfrey}},\
  }\href@noop {} {\bibfield  {journal} {\bibinfo  {journal} {Physical Review
  Letters}\ }\textbf {\bibinfo {volume} {79}},\ \bibinfo {pages} {2554}
  (\bibinfo {year} {1997})}\BibitemShut {NoStop}%
\bibitem [{\citenamefont {Isakov}\ \emph {et~al.}(2005)\citenamefont {Isakov},
  \citenamefont {Moessner},\ and\ \citenamefont {Sondhi}}]{Isakov05a}%
  \BibitemOpen
  \bibfield  {author} {\bibinfo {author} {\bibfnamefont {S.~V.}\ \bibnamefont
  {Isakov}}, \bibinfo {author} {\bibfnamefont {R.}~\bibnamefont {Moessner}}, \
  and\ \bibinfo {author} {\bibfnamefont {S.~L.}\ \bibnamefont {Sondhi}},\
  }\href {\doibase ARTN 217201} {\bibfield  {journal} {\bibinfo  {journal}
  {Physical Review Letters}\ }\textbf {\bibinfo {volume} {95}},\ \bibinfo
  {pages} {217201} (\bibinfo {year} {2005})}\BibitemShut {NoStop}%
\bibitem [{\citenamefont {Fennell}\ \emph {et~al.}(2009)\citenamefont
  {Fennell}, \citenamefont {Deen}, \citenamefont {Wildes}, \citenamefont
  {Schmalzl}, \citenamefont {Prabhakaran}, \citenamefont {Boothroyd},
  \citenamefont {Aldus}, \citenamefont {McMorrow},\ and\ \citenamefont
  {Bramwell}}]{Fennell09a}%
  \BibitemOpen
  \bibfield  {author} {\bibinfo {author} {\bibfnamefont {T.}~\bibnamefont
  {Fennell}}, \bibinfo {author} {\bibfnamefont {P.~P.}\ \bibnamefont {Deen}},
  \bibinfo {author} {\bibfnamefont {A.~R.}\ \bibnamefont {Wildes}}, \bibinfo
  {author} {\bibfnamefont {K.}~\bibnamefont {Schmalzl}}, \bibinfo {author}
  {\bibfnamefont {D.}~\bibnamefont {Prabhakaran}}, \bibinfo {author}
  {\bibfnamefont {A.~T.}\ \bibnamefont {Boothroyd}}, \bibinfo {author}
  {\bibfnamefont {R.~J.}\ \bibnamefont {Aldus}}, \bibinfo {author}
  {\bibfnamefont {D.~F.}\ \bibnamefont {McMorrow}}, \ and\ \bibinfo {author}
  {\bibfnamefont {S.~T.}\ \bibnamefont {Bramwell}},\ }\href {\doibase DOI
  10.1126/science.1177582} {\bibfield  {journal} {\bibinfo  {journal}
  {Science}\ }\textbf {\bibinfo {volume} {326}},\ \bibinfo {pages} {415}
  (\bibinfo {year} {2009})}\BibitemShut {NoStop}%
\bibitem [{\citenamefont {Henley}(2010)}]{Henley10a}%
  \BibitemOpen
  \bibfield  {author} {\bibinfo {author} {\bibfnamefont {C.}~\bibnamefont
  {Henley}},\ }\href {\doibase 10.1146/annurev-conmatphys-070909-104138}
  {\bibfield  {journal} {\bibinfo  {journal} {Annual Review of Condensed Matter
  Physics}\ }\textbf {\bibinfo {volume} {1}},\ \bibinfo {pages} {179} (\bibinfo
  {year} {2010})}\BibitemShut {NoStop}%
\bibitem [{\citenamefont {Castelnovo}\ \emph {et~al.}(2008)\citenamefont
  {Castelnovo}, \citenamefont {Moessner},\ and\ \citenamefont
  {Sondhi}}]{Castelnovo08a}%
  \BibitemOpen
  \bibfield  {author} {\bibinfo {author} {\bibfnamefont {C.}~\bibnamefont
  {Castelnovo}}, \bibinfo {author} {\bibfnamefont {R.}~\bibnamefont
  {Moessner}}, \ and\ \bibinfo {author} {\bibfnamefont {S.~L.}\ \bibnamefont
  {Sondhi}},\ }\href {\doibase DOI 10.1038/nature06433} {\bibfield  {journal}
  {\bibinfo  {journal} {Nature}\ }\textbf {\bibinfo {volume} {451}},\ \bibinfo
  {pages} {42} (\bibinfo {year} {2008})}\BibitemShut {NoStop}%
\bibitem [{\citenamefont {Khomskii}(2012)}]{Khomskii12a}%
  \BibitemOpen
  \bibfield  {author} {\bibinfo {author} {\bibfnamefont {D.}~\bibnamefont
  {Khomskii}},\ }\href@noop {} {\bibfield  {journal} {\bibinfo  {journal}
  {Nature Communications}\ }\textbf {\bibinfo {volume} {3}},\ \bibinfo {pages}
  {904} (\bibinfo {year} {2012})}\BibitemShut {NoStop}%
\bibitem [{\citenamefont {de~Leeuw}\ \emph {et~al.}(1980)\citenamefont
  {de~Leeuw}, \citenamefont {Perram},\ and\ \citenamefont
  {Smith}}]{Deleeuw80a}%
  \BibitemOpen
  \bibfield  {author} {\bibinfo {author} {\bibfnamefont {S.~W.}\ \bibnamefont
  {de~Leeuw}}, \bibinfo {author} {\bibfnamefont {J.~W.}\ \bibnamefont
  {Perram}}, \ and\ \bibinfo {author} {\bibfnamefont {E.~R.}\ \bibnamefont
  {Smith}},\ }\href@noop {} {\bibfield  {journal} {\bibinfo  {journal}
  {Proceedings of the Royal Society of London A}\ }\textbf {\bibinfo {volume}
  {373}},\ \bibinfo {pages} {27} (\bibinfo {year} {1980})}\BibitemShut
  {NoStop}%
\bibitem [{\citenamefont {den Hertog}\ and\ \citenamefont
  {Gingras}(2000)}]{Hertog00a}%
  \BibitemOpen
  \bibfield  {author} {\bibinfo {author} {\bibfnamefont {B.~C.}\ \bibnamefont
  {den Hertog}}\ and\ \bibinfo {author} {\bibfnamefont {M.~J.~P.}\ \bibnamefont
  {Gingras}},\ }\href@noop {} {\bibfield  {journal} {\bibinfo  {journal}
  {Physical Review Letters}\ }\textbf {\bibinfo {volume} {84}},\ \bibinfo
  {pages} {3430} (\bibinfo {year} {2000})}\BibitemShut {NoStop}%
\bibitem [{\citenamefont {Melko}\ and\ \citenamefont
  {Gingras}(2004)}]{Melko04a}%
  \BibitemOpen
  \bibfield  {author} {\bibinfo {author} {\bibfnamefont {R.~G.}\ \bibnamefont
  {Melko}}\ and\ \bibinfo {author} {\bibfnamefont {M.~J.~P.}\ \bibnamefont
  {Gingras}},\ }\href {\doibase DOI 10.1088/0953-8984/16/43/R02} {\bibfield
  {journal} {\bibinfo  {journal} {Journal of Physics-Condensed Matter}\
  }\textbf {\bibinfo {volume} {16}},\ \bibinfo {pages} {R1277} (\bibinfo {year}
  {2004})}\BibitemShut {NoStop}%
\bibitem [{\citenamefont {Moessner}(1998)}]{Moessner98b}%
  \BibitemOpen
  \bibfield  {author} {\bibinfo {author} {\bibfnamefont {R.}~\bibnamefont
  {Moessner}},\ }\href@noop {} {\bibfield  {journal} {\bibinfo  {journal}
  {Physical Review B}\ }\textbf {\bibinfo {volume} {57}},\ \bibinfo {pages}
  {R5587} (\bibinfo {year} {1998})}\BibitemShut {NoStop}%
\bibitem [{\citenamefont {Bramwell}\ and\ \citenamefont
  {Harris}(1998)}]{Bramwell98a}%
  \BibitemOpen
  \bibfield  {author} {\bibinfo {author} {\bibfnamefont {S.~T.}\ \bibnamefont
  {Bramwell}}\ and\ \bibinfo {author} {\bibfnamefont {M.~J.}\ \bibnamefont
  {Harris}},\ }\href@noop {} {\bibfield  {journal} {\bibinfo  {journal}
  {Journal of Physics-Condensed Matter}\ }\textbf {\bibinfo {volume} {10}},\
  \bibinfo {pages} {L215} (\bibinfo {year} {1998})}\BibitemShut {NoStop}%
\bibitem [{\citenamefont {Siddharthan}\ \emph {et~al.}(1999)\citenamefont
  {Siddharthan}, \citenamefont {Shastry}, \citenamefont {Ramirez},
  \citenamefont {Hayashi}, \citenamefont {Cava},\ and\ \citenamefont
  {Rosenkranz}}]{Siddharthan99a}%
  \BibitemOpen
  \bibfield  {author} {\bibinfo {author} {\bibfnamefont {R.}~\bibnamefont
  {Siddharthan}}, \bibinfo {author} {\bibfnamefont {B.~S.}\ \bibnamefont
  {Shastry}}, \bibinfo {author} {\bibfnamefont {A.~P.}\ \bibnamefont
  {Ramirez}}, \bibinfo {author} {\bibfnamefont {A.}~\bibnamefont {Hayashi}},
  \bibinfo {author} {\bibfnamefont {R.~J.}\ \bibnamefont {Cava}}, \ and\
  \bibinfo {author} {\bibfnamefont {S.}~\bibnamefont {Rosenkranz}},\
  }\href@noop {} {\bibfield  {journal} {\bibinfo  {journal} {Physical Review
  Letters}\ }\textbf {\bibinfo {volume} {83}},\ \bibinfo {pages} {1854}
  (\bibinfo {year} {1999})}\BibitemShut {NoStop}%
\bibitem [{\citenamefont {Melko}\ \emph {et~al.}(2001)\citenamefont {Melko},
  \citenamefont {den Hertog},\ and\ \citenamefont {Gingras}}]{Melko01a}%
  \BibitemOpen
  \bibfield  {author} {\bibinfo {author} {\bibfnamefont {R.~G.}\ \bibnamefont
  {Melko}}, \bibinfo {author} {\bibfnamefont {B.~C.}\ \bibnamefont {den
  Hertog}}, \ and\ \bibinfo {author} {\bibfnamefont {M.~J.~P.}\ \bibnamefont
  {Gingras}},\ }\href {\doibase ARTN 067203} {\bibfield  {journal} {\bibinfo
  {journal} {Physical Review Letters}\ }\textbf {\bibinfo {volume} {87}},\
  \bibinfo {pages} {067203} (\bibinfo {year} {2001})}\BibitemShut {NoStop}%
\bibitem [{\citenamefont {Higashinaka}\ \emph {et~al.}(2003)\citenamefont
  {Higashinaka}, \citenamefont {Fukazawa},\ and\ \citenamefont
  {Maeno}}]{Higashinaka03b}%
  \BibitemOpen
  \bibfield  {author} {\bibinfo {author} {\bibfnamefont {R.}~\bibnamefont
  {Higashinaka}}, \bibinfo {author} {\bibfnamefont {H.}~\bibnamefont
  {Fukazawa}}, \ and\ \bibinfo {author} {\bibfnamefont {Y.}~\bibnamefont
  {Maeno}},\ }\href {\doibase ARTN 014415} {\bibfield  {journal} {\bibinfo
  {journal} {Physical Review B}\ }\textbf {\bibinfo {volume} {68}},\ \bibinfo
  {pages} {014415} (\bibinfo {year} {2003})}\BibitemShut {NoStop}%
\bibitem [{\citenamefont {Yoshida}\ \emph {et~al.}(2004)\citenamefont
  {Yoshida}, \citenamefont {Nemoto},\ and\ \citenamefont {Wada}}]{Yoshida04a}%
  \BibitemOpen
  \bibfield  {author} {\bibinfo {author} {\bibfnamefont {S.}~\bibnamefont
  {Yoshida}}, \bibinfo {author} {\bibfnamefont {K.}~\bibnamefont {Nemoto}}, \
  and\ \bibinfo {author} {\bibfnamefont {K.}~\bibnamefont {Wada}},\ }\href
  {\doibase DOI 10.1143/JPSJ.73.1619} {\bibfield  {journal} {\bibinfo
  {journal} {Journal of the Physical Society of Japan}\ }\textbf {\bibinfo
  {volume} {73}},\ \bibinfo {pages} {1619} (\bibinfo {year}
  {2004})}\BibitemShut {NoStop}%
\bibitem [{\citenamefont {Fennell}\ \emph {et~al.}(2005)\citenamefont
  {Fennell}, \citenamefont {Petrenko}, \citenamefont {Fak}, \citenamefont
  {Gardner}, \citenamefont {Bramwell},\ and\ \citenamefont
  {Ouladdiaf}}]{Fennell05a}%
  \BibitemOpen
  \bibfield  {author} {\bibinfo {author} {\bibfnamefont {T.}~\bibnamefont
  {Fennell}}, \bibinfo {author} {\bibfnamefont {O.~A.}\ \bibnamefont
  {Petrenko}}, \bibinfo {author} {\bibfnamefont {B.}~\bibnamefont {Fak}},
  \bibinfo {author} {\bibfnamefont {J.~S.}\ \bibnamefont {Gardner}}, \bibinfo
  {author} {\bibfnamefont {S.~T.}\ \bibnamefont {Bramwell}}, \ and\ \bibinfo
  {author} {\bibfnamefont {B.}~\bibnamefont {Ouladdiaf}},\ }\href {\doibase
  ARTN 224411} {\bibfield  {journal} {\bibinfo  {journal} {Physical Review B}\
  }\textbf {\bibinfo {volume} {72}},\ \bibinfo {pages} {224411} (\bibinfo
  {year} {2005})}\BibitemShut {NoStop}%
\bibitem [{\citenamefont {Ruff}\ \emph
  {et~al.}(2010{\natexlab{a}})\citenamefont {Ruff}, \citenamefont {Gaulin},
  \citenamefont {Rule},\ and\ \citenamefont {Gardner}}]{Ruff10a}%
  \BibitemOpen
  \bibfield  {author} {\bibinfo {author} {\bibfnamefont {J.~P.~C.}\
  \bibnamefont {Ruff}}, \bibinfo {author} {\bibfnamefont {B.~D.}\ \bibnamefont
  {Gaulin}}, \bibinfo {author} {\bibfnamefont {K.~C.}\ \bibnamefont {Rule}}, \
  and\ \bibinfo {author} {\bibfnamefont {J.~S.}\ \bibnamefont {Gardner}},\
  }\href {\doibase 10.1103/PhysRevB.82.100401} {\bibfield  {journal} {\bibinfo
  {journal} {Phys. Rev. B}\ }\textbf {\bibinfo {volume} {82}},\ \bibinfo
  {pages} {100401} (\bibinfo {year} {2010}{\natexlab{a}})}\BibitemShut
  {NoStop}%
\bibitem [{\citenamefont {Sazonov}\ \emph {et~al.}(2012)\citenamefont
  {Sazonov}, \citenamefont {Gukasov}, \citenamefont {Mirebeau},\ and\
  \citenamefont {Bonville}}]{Sazonov12a}%
  \BibitemOpen
  \bibfield  {author} {\bibinfo {author} {\bibfnamefont {A.~P.}\ \bibnamefont
  {Sazonov}}, \bibinfo {author} {\bibfnamefont {A.}~\bibnamefont {Gukasov}},
  \bibinfo {author} {\bibfnamefont {I.}~\bibnamefont {Mirebeau}}, \ and\
  \bibinfo {author} {\bibfnamefont {P.}~\bibnamefont {Bonville}},\ }\href
  {\doibase 10.1103/PhysRevB.85.214420} {\bibfield  {journal} {\bibinfo
  {journal} {Phys. Rev. B}\ }\textbf {\bibinfo {volume} {85}},\ \bibinfo
  {pages} {214420} (\bibinfo {year} {2012})}\BibitemShut {NoStop}%
\bibitem [{\citenamefont {Gingras}\ \emph {et~al.}(2000)\citenamefont
  {Gingras}, \citenamefont {den Hertog}, \citenamefont {Faucher}, \citenamefont
  {Gardner}, \citenamefont {Dunsiger}, \citenamefont {Chang}, \citenamefont
  {Gaulin}, \citenamefont {Raju},\ and\ \citenamefont {Greedan}}]{Gingras00a}%
  \BibitemOpen
  \bibfield  {author} {\bibinfo {author} {\bibfnamefont {M.~J.~P.}\
  \bibnamefont {Gingras}}, \bibinfo {author} {\bibfnamefont {B.~C.}\
  \bibnamefont {den Hertog}}, \bibinfo {author} {\bibfnamefont
  {M.}~\bibnamefont {Faucher}}, \bibinfo {author} {\bibfnamefont {J.~S.}\
  \bibnamefont {Gardner}}, \bibinfo {author} {\bibfnamefont {S.~R.}\
  \bibnamefont {Dunsiger}}, \bibinfo {author} {\bibfnamefont {L.~J.}\
  \bibnamefont {Chang}}, \bibinfo {author} {\bibfnamefont {B.~D.}\ \bibnamefont
  {Gaulin}}, \bibinfo {author} {\bibfnamefont {N.~P.}\ \bibnamefont {Raju}}, \
  and\ \bibinfo {author} {\bibfnamefont {J.~E.}\ \bibnamefont {Greedan}},\
  }\href@noop {} {\bibfield  {journal} {\bibinfo  {journal} {Physical Review
  B}\ }\textbf {\bibinfo {volume} {62}},\ \bibinfo {pages} {6496} (\bibinfo
  {year} {2000})}\BibitemShut {NoStop}%
\bibitem [{\citenamefont {Cao}\ \emph {et~al.}(2009)\citenamefont {Cao},
  \citenamefont {Gukasov}, \citenamefont {Mirebeau}, \citenamefont {Bonville},
  \citenamefont {Decorse},\ and\ \citenamefont {Dhalenne}}]{Cao09b}%
  \BibitemOpen
  \bibfield  {author} {\bibinfo {author} {\bibfnamefont {H.}~\bibnamefont
  {Cao}}, \bibinfo {author} {\bibfnamefont {A.}~\bibnamefont {Gukasov}},
  \bibinfo {author} {\bibfnamefont {I.}~\bibnamefont {Mirebeau}}, \bibinfo
  {author} {\bibfnamefont {P.}~\bibnamefont {Bonville}}, \bibinfo {author}
  {\bibfnamefont {C.}~\bibnamefont {Decorse}}, \ and\ \bibinfo {author}
  {\bibfnamefont {G.}~\bibnamefont {Dhalenne}},\ }\href {\doibase
  10.1103/PhysRevLett.103.056402} {\bibfield  {journal} {\bibinfo  {journal}
  {Phys. Rev. Lett.}\ }\textbf {\bibinfo {volume} {103}},\ \bibinfo {pages}
  {056402} (\bibinfo {year} {2009})}\BibitemShut {NoStop}%
\bibitem [{\citenamefont {Fennell}\ \emph {et~al.}(2012)\citenamefont
  {Fennell}, \citenamefont {Kenzelmann}, \citenamefont {Roessli}, \citenamefont
  {Haas},\ and\ \citenamefont {Cava}}]{Fennell12a}%
  \BibitemOpen
  \bibfield  {author} {\bibinfo {author} {\bibfnamefont {T.}~\bibnamefont
  {Fennell}}, \bibinfo {author} {\bibfnamefont {M.}~\bibnamefont {Kenzelmann}},
  \bibinfo {author} {\bibfnamefont {B.}~\bibnamefont {Roessli}}, \bibinfo
  {author} {\bibfnamefont {M.~K.}\ \bibnamefont {Haas}}, \ and\ \bibinfo
  {author} {\bibfnamefont {R.~J.}\ \bibnamefont {Cava}},\ }\href {\doibase
  10.1103/PhysRevLett.109.017201} {\bibfield  {journal} {\bibinfo  {journal}
  {Phys. Rev. Lett.}\ }\textbf {\bibinfo {volume} {109}},\ \bibinfo {pages}
  {017201} (\bibinfo {year} {2012})}\BibitemShut {NoStop}%
\bibitem [{\citenamefont {Petit}\ \emph {et~al.}(2012)\citenamefont {Petit},
  \citenamefont {Bonville}, \citenamefont {Robert}, \citenamefont {Decorse},\
  and\ \citenamefont {Mirebeau}}]{Petit12a}%
  \BibitemOpen
  \bibfield  {author} {\bibinfo {author} {\bibfnamefont {S.}~\bibnamefont
  {Petit}}, \bibinfo {author} {\bibfnamefont {P.}~\bibnamefont {Bonville}},
  \bibinfo {author} {\bibfnamefont {J.}~\bibnamefont {Robert}}, \bibinfo
  {author} {\bibfnamefont {C.}~\bibnamefont {Decorse}}, \ and\ \bibinfo
  {author} {\bibfnamefont {I.}~\bibnamefont {Mirebeau}},\ }\href {\doibase
  10.1103/PhysRevB.86.174403} {\bibfield  {journal} {\bibinfo  {journal} {Phys.
  Rev. B}\ }\textbf {\bibinfo {volume} {86}},\ \bibinfo {pages} {174403}
  (\bibinfo {year} {2012})}\BibitemShut {NoStop}%
\bibitem [{\citenamefont {Guitteny}\ \emph {et~al.}(2013)\citenamefont
  {Guitteny}, \citenamefont {Robert}, \citenamefont {Bonville}, \citenamefont
  {Ollivier}, \citenamefont {Decorse}, \citenamefont {Steffens}, \citenamefont
  {Boehm}, \citenamefont {Mutka}, \citenamefont {Mirebeau},\ and\ \citenamefont
  {Petit}}]{Guitteny13a}%
  \BibitemOpen
  \bibfield  {author} {\bibinfo {author} {\bibfnamefont {S.}~\bibnamefont
  {Guitteny}}, \bibinfo {author} {\bibfnamefont {J.}~\bibnamefont {Robert}},
  \bibinfo {author} {\bibfnamefont {P.}~\bibnamefont {Bonville}}, \bibinfo
  {author} {\bibfnamefont {J.}~\bibnamefont {Ollivier}}, \bibinfo {author}
  {\bibfnamefont {C.}~\bibnamefont {Decorse}}, \bibinfo {author} {\bibfnamefont
  {P.}~\bibnamefont {Steffens}}, \bibinfo {author} {\bibfnamefont
  {M.}~\bibnamefont {Boehm}}, \bibinfo {author} {\bibfnamefont
  {H.}~\bibnamefont {Mutka}}, \bibinfo {author} {\bibfnamefont
  {I.}~\bibnamefont {Mirebeau}}, \ and\ \bibinfo {author} {\bibfnamefont
  {S.}~\bibnamefont {Petit}},\ }\href {\doibase 10.1103/PhysRevLett.111.087201}
  {\bibfield  {journal} {\bibinfo  {journal} {Phys. Rev. Lett.}\ }\textbf
  {\bibinfo {volume} {111}},\ \bibinfo {pages} {087201} (\bibinfo {year}
  {2013})}\BibitemShut {NoStop}%
\bibitem [{\citenamefont {Aleksandrov}\ \emph {et~al.}(1985)\citenamefont
  {Aleksandrov}, \citenamefont {Lidskii}, \citenamefont {Mamsurova},
  \citenamefont {Neigauz}, \citenamefont {Pigal'skii}, \citenamefont {Pukhov},
  \citenamefont {Trusevich},\ and\ \citenamefont
  {Chcherbakova}}]{Aleksandrov85a}%
  \BibitemOpen
  \bibfield  {author} {\bibinfo {author} {\bibfnamefont {I.}~\bibnamefont
  {Aleksandrov}}, \bibinfo {author} {\bibfnamefont {B.}~\bibnamefont
  {Lidskii}}, \bibinfo {author} {\bibfnamefont {L.}~\bibnamefont {Mamsurova}},
  \bibinfo {author} {\bibfnamefont {M.}~\bibnamefont {Neigauz}}, \bibinfo
  {author} {\bibfnamefont {K.}~\bibnamefont {Pigal'skii}}, \bibinfo {author}
  {\bibfnamefont {K.}~\bibnamefont {Pukhov}}, \bibinfo {author} {\bibfnamefont
  {N.}~\bibnamefont {Trusevich}}, \ and\ \bibinfo {author} {\bibfnamefont
  {L.}~\bibnamefont {Chcherbakova}},\ }\href@noop {} {\bibfield  {journal}
  {\bibinfo  {journal} {Journal of Experimental and Theoretical Physics}\
  }\textbf {\bibinfo {volume} {89}},\ \bibinfo {pages} {2230} (\bibinfo {year}
  {1985})}\BibitemShut {NoStop}%
\bibitem [{\citenamefont {Mamsurova}\ \emph {et~al.}(1986)\citenamefont
  {Mamsurova}, \citenamefont {Pigal'skii},\ and\ \citenamefont
  {Pukhov}}]{Mamsurova86a}%
  \BibitemOpen
  \bibfield  {author} {\bibinfo {author} {\bibfnamefont {L.}~\bibnamefont
  {Mamsurova}}, \bibinfo {author} {\bibfnamefont {K.}~\bibnamefont
  {Pigal'skii}}, \ and\ \bibinfo {author} {\bibfnamefont {K.}~\bibnamefont
  {Pukhov}},\ }\href@noop {} {\bibfield  {journal} {\bibinfo  {journal}
  {Journal of Experimental and Theoretical Physics}\ }\textbf {\bibinfo
  {volume} {43}},\ \bibinfo {pages} {584} (\bibinfo {year} {1986})}\BibitemShut
  {NoStop}%
\bibitem [{\citenamefont {Ruff}\ \emph
  {et~al.}(2010{\natexlab{b}})\citenamefont {Ruff}, \citenamefont {Islam},
  \citenamefont {Clancy}, \citenamefont {Ross}, \citenamefont {Nojiri},
  \citenamefont {Matsuda}, \citenamefont {Dabkowska}, \citenamefont
  {Dabkowski},\ and\ \citenamefont {Gaulin}}]{Ruff10b}%
  \BibitemOpen
  \bibfield  {author} {\bibinfo {author} {\bibfnamefont {J.~P.~C.}\
  \bibnamefont {Ruff}}, \bibinfo {author} {\bibfnamefont {Z.}~\bibnamefont
  {Islam}}, \bibinfo {author} {\bibfnamefont {J.~P.}\ \bibnamefont {Clancy}},
  \bibinfo {author} {\bibfnamefont {K.~A.}\ \bibnamefont {Ross}}, \bibinfo
  {author} {\bibfnamefont {H.}~\bibnamefont {Nojiri}}, \bibinfo {author}
  {\bibfnamefont {Y.~H.}\ \bibnamefont {Matsuda}}, \bibinfo {author}
  {\bibfnamefont {H.~A.}\ \bibnamefont {Dabkowska}}, \bibinfo {author}
  {\bibfnamefont {A.~D.}\ \bibnamefont {Dabkowski}}, \ and\ \bibinfo {author}
  {\bibfnamefont {B.~D.}\ \bibnamefont {Gaulin}},\ }\href {\doibase
  10.1103/PhysRevLett.105.077203} {\bibfield  {journal} {\bibinfo  {journal}
  {Phys. Rev. Lett.}\ }\textbf {\bibinfo {volume} {105}},\ \bibinfo {pages}
  {077203} (\bibinfo {year} {2010}{\natexlab{b}})}\BibitemShut {NoStop}%
\bibitem [{\citenamefont {Fennell}\ \emph {et~al.}(2014)\citenamefont
  {Fennell}, \citenamefont {Kenzelmann}, \citenamefont {Roessli}, \citenamefont
  {Mutka}, \citenamefont {Ollivier}, \citenamefont {Ruminy}, \citenamefont
  {Stuhr}, \citenamefont {Zaharko}, \citenamefont {Bovo}, \citenamefont
  {Cervellino}, \citenamefont {Haas},\ and\ \citenamefont {Cava}}]{Fennell14a}%
  \BibitemOpen
  \bibfield  {author} {\bibinfo {author} {\bibfnamefont {T.}~\bibnamefont
  {Fennell}}, \bibinfo {author} {\bibfnamefont {M.}~\bibnamefont {Kenzelmann}},
  \bibinfo {author} {\bibfnamefont {B.}~\bibnamefont {Roessli}}, \bibinfo
  {author} {\bibfnamefont {H.}~\bibnamefont {Mutka}}, \bibinfo {author}
  {\bibfnamefont {J.}~\bibnamefont {Ollivier}}, \bibinfo {author}
  {\bibfnamefont {M.}~\bibnamefont {Ruminy}}, \bibinfo {author} {\bibfnamefont
  {U.}~\bibnamefont {Stuhr}}, \bibinfo {author} {\bibfnamefont
  {O.}~\bibnamefont {Zaharko}}, \bibinfo {author} {\bibfnamefont
  {L.}~\bibnamefont {Bovo}}, \bibinfo {author} {\bibfnamefont {A.}~\bibnamefont
  {Cervellino}}, \bibinfo {author} {\bibfnamefont {M.~K.}\ \bibnamefont
  {Haas}}, \ and\ \bibinfo {author} {\bibfnamefont {R.~J.}\ \bibnamefont
  {Cava}},\ }\href {\doibase 10.1103/PhysRevLett.112.017203} {\bibfield
  {journal} {\bibinfo  {journal} {Phys. Rev. Lett.}\ }\textbf {\bibinfo
  {volume} {112}},\ \bibinfo {pages} {017203} (\bibinfo {year}
  {2014})}\BibitemShut {NoStop}%
\bibitem [{\citenamefont {Bonville}\ \emph {et~al.}(2014)\citenamefont
  {Bonville}, \citenamefont {Gukasov}, \citenamefont {Mirebeau},\ and\
  \citenamefont {Petit}}]{Bonville14a}%
  \BibitemOpen
  \bibfield  {author} {\bibinfo {author} {\bibfnamefont {P.}~\bibnamefont
  {Bonville}}, \bibinfo {author} {\bibfnamefont {A.}~\bibnamefont {Gukasov}},
  \bibinfo {author} {\bibfnamefont {I.}~\bibnamefont {Mirebeau}}, \ and\
  \bibinfo {author} {\bibfnamefont {S.}~\bibnamefont {Petit}},\ }\href
  {\doibase 10.1103/PhysRevB.89.085115} {\bibfield  {journal} {\bibinfo
  {journal} {Phys. Rev. B}\ }\textbf {\bibinfo {volume} {89}},\ \bibinfo
  {pages} {085115} (\bibinfo {year} {2014})}\BibitemShut {NoStop}%
\bibitem [{\citenamefont {Taniguchi}\ \emph {et~al.}(2013)\citenamefont
  {Taniguchi}, \citenamefont {Kadowaki}, \citenamefont {Takatsu}, \citenamefont
  {F\aa{}k}, \citenamefont {Ollivier}, \citenamefont {Yamazaki}, \citenamefont
  {Sato}, \citenamefont {Yoshizawa}, \citenamefont {Shimura}, \citenamefont
  {Sakakibara}, \citenamefont {Hong}, \citenamefont {Goto}, \citenamefont
  {Yaraskavitch},\ and\ \citenamefont {Kycia}}]{Taniguchi13a}%
  \BibitemOpen
  \bibfield  {author} {\bibinfo {author} {\bibfnamefont {T.}~\bibnamefont
  {Taniguchi}}, \bibinfo {author} {\bibfnamefont {H.}~\bibnamefont {Kadowaki}},
  \bibinfo {author} {\bibfnamefont {H.}~\bibnamefont {Takatsu}}, \bibinfo
  {author} {\bibfnamefont {B.}~\bibnamefont {F\aa{}k}}, \bibinfo {author}
  {\bibfnamefont {J.}~\bibnamefont {Ollivier}}, \bibinfo {author}
  {\bibfnamefont {T.}~\bibnamefont {Yamazaki}}, \bibinfo {author}
  {\bibfnamefont {T.~J.}\ \bibnamefont {Sato}}, \bibinfo {author}
  {\bibfnamefont {H.}~\bibnamefont {Yoshizawa}}, \bibinfo {author}
  {\bibfnamefont {Y.}~\bibnamefont {Shimura}}, \bibinfo {author} {\bibfnamefont
  {T.}~\bibnamefont {Sakakibara}}, \bibinfo {author} {\bibfnamefont
  {T.}~\bibnamefont {Hong}}, \bibinfo {author} {\bibfnamefont {K.}~\bibnamefont
  {Goto}}, \bibinfo {author} {\bibfnamefont {L.~R.}\ \bibnamefont
  {Yaraskavitch}}, \ and\ \bibinfo {author} {\bibfnamefont {J.~B.}\
  \bibnamefont {Kycia}},\ }\href {\doibase 10.1103/PhysRevB.87.060408}
  {\bibfield  {journal} {\bibinfo  {journal} {Phys. Rev. B}\ }\textbf {\bibinfo
  {volume} {87}},\ \bibinfo {pages} {060408} (\bibinfo {year}
  {2013})}\BibitemShut {NoStop}%
\bibitem [{\citenamefont {Gardner}\ \emph {et~al.}(1999)\citenamefont
  {Gardner}, \citenamefont {Dunsiger}, \citenamefont {Gaulin}, \citenamefont
  {Gingras}, \citenamefont {Greedan}, \citenamefont {Kiefl}, \citenamefont
  {Lumsden}, \citenamefont {MacFarlane}, \citenamefont {Raju}, \citenamefont
  {Sonier}, \citenamefont {Swainson},\ and\ \citenamefont {Tun}}]{Gardner99a}%
  \BibitemOpen
  \bibfield  {author} {\bibinfo {author} {\bibfnamefont {J.~S.}\ \bibnamefont
  {Gardner}}, \bibinfo {author} {\bibfnamefont {S.~R.}\ \bibnamefont
  {Dunsiger}}, \bibinfo {author} {\bibfnamefont {B.~D.}\ \bibnamefont
  {Gaulin}}, \bibinfo {author} {\bibfnamefont {M.~J.~P.}\ \bibnamefont
  {Gingras}}, \bibinfo {author} {\bibfnamefont {J.~E.}\ \bibnamefont
  {Greedan}}, \bibinfo {author} {\bibfnamefont {R.~F.}\ \bibnamefont {Kiefl}},
  \bibinfo {author} {\bibfnamefont {M.~D.}\ \bibnamefont {Lumsden}}, \bibinfo
  {author} {\bibfnamefont {W.~A.}\ \bibnamefont {MacFarlane}}, \bibinfo
  {author} {\bibfnamefont {N.~P.}\ \bibnamefont {Raju}}, \bibinfo {author}
  {\bibfnamefont {J.~E.}\ \bibnamefont {Sonier}}, \bibinfo {author}
  {\bibfnamefont {I.}~\bibnamefont {Swainson}}, \ and\ \bibinfo {author}
  {\bibfnamefont {Z.}~\bibnamefont {Tun}},\ }\href@noop {} {\bibfield
  {journal} {\bibinfo  {journal} {Physical Review Letters}\ }\textbf {\bibinfo
  {volume} {82}},\ \bibinfo {pages} {1012} (\bibinfo {year}
  {1999})}\BibitemShut {NoStop}%
\bibitem [{\citenamefont {Gardner}\ \emph {et~al.}(2003)\citenamefont
  {Gardner}, \citenamefont {Keren}, \citenamefont {Ehlers}, \citenamefont
  {Stock}, \citenamefont {Segal}, \citenamefont {Roper}, \citenamefont {Fak},
  \citenamefont {Stone}, \citenamefont {Hammar}, \citenamefont {Reich},\ and\
  \citenamefont {Gaulin}}]{Gardner03a}%
  \BibitemOpen
  \bibfield  {author} {\bibinfo {author} {\bibfnamefont {J.~S.}\ \bibnamefont
  {Gardner}}, \bibinfo {author} {\bibfnamefont {A.}~\bibnamefont {Keren}},
  \bibinfo {author} {\bibfnamefont {G.}~\bibnamefont {Ehlers}}, \bibinfo
  {author} {\bibfnamefont {C.}~\bibnamefont {Stock}}, \bibinfo {author}
  {\bibfnamefont {E.}~\bibnamefont {Segal}}, \bibinfo {author} {\bibfnamefont
  {J.~M.}\ \bibnamefont {Roper}}, \bibinfo {author} {\bibfnamefont
  {B.}~\bibnamefont {Fak}}, \bibinfo {author} {\bibfnamefont {M.~B.}\
  \bibnamefont {Stone}}, \bibinfo {author} {\bibfnamefont {P.~R.}\ \bibnamefont
  {Hammar}}, \bibinfo {author} {\bibfnamefont {D.~H.}\ \bibnamefont {Reich}}, \
  and\ \bibinfo {author} {\bibfnamefont {B.~D.}\ \bibnamefont {Gaulin}},\
  }\href {\doibase ARTN 180401} {\bibfield  {journal} {\bibinfo  {journal}
  {Physical Review B}\ }\textbf {\bibinfo {volume} {68}},\ \bibinfo {pages}
  {180401} (\bibinfo {year} {2003})}\BibitemShut {NoStop}%
\bibitem [{\citenamefont {Bonville}\ \emph {et~al.}(2011)\citenamefont
  {Bonville}, \citenamefont {Mirebeau}, \citenamefont {Gukasov}, \citenamefont
  {Petit},\ and\ \citenamefont {Robert}}]{Bonville11a}%
  \BibitemOpen
  \bibfield  {author} {\bibinfo {author} {\bibfnamefont {P.}~\bibnamefont
  {Bonville}}, \bibinfo {author} {\bibfnamefont {I.}~\bibnamefont {Mirebeau}},
  \bibinfo {author} {\bibfnamefont {A.}~\bibnamefont {Gukasov}}, \bibinfo
  {author} {\bibfnamefont {S.}~\bibnamefont {Petit}}, \ and\ \bibinfo {author}
  {\bibfnamefont {J.}~\bibnamefont {Robert}},\ }\href {\doibase
  10.1103/PhysRevB.84.184409} {\bibfield  {journal} {\bibinfo  {journal} {Phys.
  Rev. B}\ }\textbf {\bibinfo {volume} {84}},\ \bibinfo {pages} {184409}
  (\bibinfo {year} {2011})}\BibitemShut {NoStop}%
\bibitem [{\citenamefont {Luo}\ \emph {et~al.}(2001)\citenamefont {Luo},
  \citenamefont {Hess},\ and\ \citenamefont {Corruccini}}]{Luo01a}%
  \BibitemOpen
  \bibfield  {author} {\bibinfo {author} {\bibfnamefont {G.}~\bibnamefont
  {Luo}}, \bibinfo {author} {\bibfnamefont {S.}~\bibnamefont {Hess}}, \ and\
  \bibinfo {author} {\bibfnamefont {L.}~\bibnamefont {Corruccini}},\
  }\href@noop {} {\bibfield  {journal} {\bibinfo  {journal} {Physics Letters
  A}\ }\textbf {\bibinfo {volume} {291}},\ \bibinfo {pages} {306} (\bibinfo
  {year} {2001})}\BibitemShut {NoStop}%
\bibitem [{\citenamefont {Yasui}\ \emph {et~al.}(2002)\citenamefont {Yasui},
  \citenamefont {Kanada}, \citenamefont {Ito}, \citenamefont {Harashina},
  \citenamefont {Sato}, \citenamefont {Okumura}, \citenamefont {Kakurai},\ and\
  \citenamefont {Kadowaki}}]{Yasui02a}%
  \BibitemOpen
  \bibfield  {author} {\bibinfo {author} {\bibfnamefont {Y.}~\bibnamefont
  {Yasui}}, \bibinfo {author} {\bibfnamefont {M.}~\bibnamefont {Kanada}},
  \bibinfo {author} {\bibfnamefont {M.}~\bibnamefont {Ito}}, \bibinfo {author}
  {\bibfnamefont {H.}~\bibnamefont {Harashina}}, \bibinfo {author}
  {\bibfnamefont {M.}~\bibnamefont {Sato}}, \bibinfo {author} {\bibfnamefont
  {H.}~\bibnamefont {Okumura}}, \bibinfo {author} {\bibfnamefont
  {K.}~\bibnamefont {Kakurai}}, \ and\ \bibinfo {author} {\bibfnamefont
  {H.}~\bibnamefont {Kadowaki}},\ }\href {\doibase DOI 10.1143/JPSJ.71.599}
  {\bibfield  {journal} {\bibinfo  {journal} {Journal of the Physical Society
  of Japan}\ }\textbf {\bibinfo {volume} {71}},\ \bibinfo {pages} {599}
  (\bibinfo {year} {2002})}\BibitemShut {NoStop}%
\bibitem [{\citenamefont {Fritsch}\ \emph {et~al.}(2014)\citenamefont
  {Fritsch}, \citenamefont {Kermarrec}, \citenamefont {Ross}, \citenamefont
  {Qiu}, \citenamefont {Copley}, \citenamefont {Pomaranski}, \citenamefont
  {Kycia}, \citenamefont {Dabkowska},\ and\ \citenamefont
  {Gaulin}}]{Fritsch14a}%
  \BibitemOpen
  \bibfield  {author} {\bibinfo {author} {\bibfnamefont {K.}~\bibnamefont
  {Fritsch}}, \bibinfo {author} {\bibfnamefont {E.}~\bibnamefont {Kermarrec}},
  \bibinfo {author} {\bibfnamefont {K.~A.}\ \bibnamefont {Ross}}, \bibinfo
  {author} {\bibfnamefont {Y.}~\bibnamefont {Qiu}}, \bibinfo {author}
  {\bibfnamefont {J.~R.~D.}\ \bibnamefont {Copley}}, \bibinfo {author}
  {\bibfnamefont {D.}~\bibnamefont {Pomaranski}}, \bibinfo {author}
  {\bibfnamefont {J.~B.}\ \bibnamefont {Kycia}}, \bibinfo {author}
  {\bibfnamefont {H.~A.}\ \bibnamefont {Dabkowska}}, \ and\ \bibinfo {author}
  {\bibfnamefont {B.~D.}\ \bibnamefont {Gaulin}},\ }\href {\doibase
  10.1103/PhysRevB.90.014429} {\bibfield  {journal} {\bibinfo  {journal} {Phys.
  Rev. B}\ }\textbf {\bibinfo {volume} {90}},\ \bibinfo {pages} {014429}
  (\bibinfo {year} {2014})}\BibitemShut {NoStop}%
\bibitem [{\citenamefont {Yasui}\ \emph {et~al.}(2001)\citenamefont {Yasui},
  \citenamefont {Kanada}, \citenamefont {Ito}, \citenamefont {Harashina},
  \citenamefont {Sato}, \citenamefont {Okumura},\ and\ \citenamefont
  {Kakurai}}]{Yasui01a}%
  \BibitemOpen
  \bibfield  {author} {\bibinfo {author} {\bibfnamefont {Y.}~\bibnamefont
  {Yasui}}, \bibinfo {author} {\bibfnamefont {M.}~\bibnamefont {Kanada}},
  \bibinfo {author} {\bibfnamefont {M.}~\bibnamefont {Ito}}, \bibinfo {author}
  {\bibfnamefont {H.}~\bibnamefont {Harashina}}, \bibinfo {author}
  {\bibfnamefont {M.}~\bibnamefont {Sato}}, \bibinfo {author} {\bibfnamefont
  {H.}~\bibnamefont {Okumura}}, \ and\ \bibinfo {author} {\bibfnamefont
  {K.}~\bibnamefont {Kakurai}},\ }\href {\doibase
  http://dx.doi.org/10.1016/S0022-3697(00)00160-8} {\bibfield  {journal}
  {\bibinfo  {journal} {Journal of Physics and Chemistry of Solids}\ }\textbf
  {\bibinfo {volume} {62}},\ \bibinfo {pages} {343 } (\bibinfo {year}
  {2001})}\BibitemShut {NoStop}%
\bibitem [{\citenamefont {Ross}\ \emph {et~al.}(2011)\citenamefont {Ross},
  \citenamefont {Savary}, \citenamefont {Gaulin},\ and\ \citenamefont
  {Balents}}]{Ross11a}%
  \BibitemOpen
  \bibfield  {author} {\bibinfo {author} {\bibfnamefont {K.~A.}\ \bibnamefont
  {Ross}}, \bibinfo {author} {\bibfnamefont {L.}~\bibnamefont {Savary}},
  \bibinfo {author} {\bibfnamefont {B.~D.}\ \bibnamefont {Gaulin}}, \ and\
  \bibinfo {author} {\bibfnamefont {L.}~\bibnamefont {Balents}},\ }\href@noop
  {} {\bibfield  {journal} {\bibinfo  {journal} {Physical Review X}\ }\textbf
  {\bibinfo {volume} {1}},\ \bibinfo {pages} {021002} (\bibinfo {year}
  {2011})}\BibitemShut {NoStop}%
\bibitem [{\citenamefont {Yaouanc}\ \emph {et~al.}(2011)\citenamefont
  {Yaouanc}, \citenamefont {Dalmas~de R\'eotier}, \citenamefont {Marin},\ and\
  \citenamefont {Glazkov}}]{Yaouanc11a}%
  \BibitemOpen
  \bibfield  {author} {\bibinfo {author} {\bibfnamefont {A.}~\bibnamefont
  {Yaouanc}}, \bibinfo {author} {\bibfnamefont {P.}~\bibnamefont {Dalmas~de
  R\'eotier}}, \bibinfo {author} {\bibfnamefont {C.}~\bibnamefont {Marin}}, \
  and\ \bibinfo {author} {\bibfnamefont {V.}~\bibnamefont {Glazkov}},\ }\href
  {\doibase 10.1103/PhysRevB.84.172408} {\bibfield  {journal} {\bibinfo
  {journal} {Phys. Rev. B}\ }\textbf {\bibinfo {volume} {84}},\ \bibinfo
  {pages} {172408} (\bibinfo {year} {2011})}\BibitemShut {NoStop}%
\bibitem [{\citenamefont {Ross}\ \emph {et~al.}(2012)\citenamefont {Ross},
  \citenamefont {Proffen}, \citenamefont {Dabkowska}, \citenamefont {Quilliam},
  \citenamefont {Yaraskavitch}, \citenamefont {Kycia},\ and\ \citenamefont
  {Gaulin}}]{Ross12a}%
  \BibitemOpen
  \bibfield  {author} {\bibinfo {author} {\bibfnamefont {K.~A.}\ \bibnamefont
  {Ross}}, \bibinfo {author} {\bibfnamefont {T.}~\bibnamefont {Proffen}},
  \bibinfo {author} {\bibfnamefont {H.~A.}\ \bibnamefont {Dabkowska}}, \bibinfo
  {author} {\bibfnamefont {J.~A.}\ \bibnamefont {Quilliam}}, \bibinfo {author}
  {\bibfnamefont {L.~R.}\ \bibnamefont {Yaraskavitch}}, \bibinfo {author}
  {\bibfnamefont {J.~B.}\ \bibnamefont {Kycia}}, \ and\ \bibinfo {author}
  {\bibfnamefont {B.~D.}\ \bibnamefont {Gaulin}},\ }\href {\doibase
  10.1103/PhysRevB.86.174424} {\bibfield  {journal} {\bibinfo  {journal} {Phys.
  Rev. B}\ }\textbf {\bibinfo {volume} {86}},\ \bibinfo {pages} {174424}
  (\bibinfo {year} {2012})}\BibitemShut {NoStop}%
\bibitem [{\citenamefont {D'Ortenzio}\ \emph {et~al.}(2013)\citenamefont
  {D'Ortenzio}, \citenamefont {Dabkowska}, \citenamefont {Dunsiger},
  \citenamefont {Gaulin}, \citenamefont {Gingras}, \citenamefont {Goko},
  \citenamefont {Kycia}, \citenamefont {Liu}, \citenamefont {Medina},
  \citenamefont {Munsie}, \citenamefont {Pomaranski}, \citenamefont {Ross},
  \citenamefont {Uemura}, \citenamefont {Williams},\ and\ \citenamefont
  {Luke}}]{Ortenzio13a}%
  \BibitemOpen
  \bibfield  {author} {\bibinfo {author} {\bibfnamefont {R.~M.}\ \bibnamefont
  {D'Ortenzio}}, \bibinfo {author} {\bibfnamefont {H.~A.}\ \bibnamefont
  {Dabkowska}}, \bibinfo {author} {\bibfnamefont {S.~R.}\ \bibnamefont
  {Dunsiger}}, \bibinfo {author} {\bibfnamefont {B.~D.}\ \bibnamefont
  {Gaulin}}, \bibinfo {author} {\bibfnamefont {M.~J.~P.}\ \bibnamefont
  {Gingras}}, \bibinfo {author} {\bibfnamefont {T.}~\bibnamefont {Goko}},
  \bibinfo {author} {\bibfnamefont {J.~B.}\ \bibnamefont {Kycia}}, \bibinfo
  {author} {\bibfnamefont {L.}~\bibnamefont {Liu}}, \bibinfo {author}
  {\bibfnamefont {T.}~\bibnamefont {Medina}}, \bibinfo {author} {\bibfnamefont
  {T.~J.}\ \bibnamefont {Munsie}}, \bibinfo {author} {\bibfnamefont
  {D.}~\bibnamefont {Pomaranski}}, \bibinfo {author} {\bibfnamefont {K.~A.}\
  \bibnamefont {Ross}}, \bibinfo {author} {\bibfnamefont {Y.~J.}\ \bibnamefont
  {Uemura}}, \bibinfo {author} {\bibfnamefont {T.~J.}\ \bibnamefont
  {Williams}}, \ and\ \bibinfo {author} {\bibfnamefont {G.~M.}\ \bibnamefont
  {Luke}},\ }\href {\doibase 10.1103/PhysRevB.88.134428} {\bibfield  {journal}
  {\bibinfo  {journal} {Phys. Rev. B}\ }\textbf {\bibinfo {volume} {88}},\
  \bibinfo {pages} {134428} (\bibinfo {year} {2013})}\BibitemShut {NoStop}%
\bibitem [{\citenamefont {Chang}\ \emph {et~al.}(2014)\citenamefont {Chang},
  \citenamefont {Lees}, \citenamefont {Watanabe}, \citenamefont {Hillier},
  \citenamefont {Yasui},\ and\ \citenamefont {Onoda}}]{Chang14a}%
  \BibitemOpen
  \bibfield  {author} {\bibinfo {author} {\bibfnamefont {L.-J.}\ \bibnamefont
  {Chang}}, \bibinfo {author} {\bibfnamefont {M.~R.}\ \bibnamefont {Lees}},
  \bibinfo {author} {\bibfnamefont {I.}~\bibnamefont {Watanabe}}, \bibinfo
  {author} {\bibfnamefont {A.~D.}\ \bibnamefont {Hillier}}, \bibinfo {author}
  {\bibfnamefont {Y.}~\bibnamefont {Yasui}}, \ and\ \bibinfo {author}
  {\bibfnamefont {S.}~\bibnamefont {Onoda}},\ }\href {\doibase
  10.1103/PhysRevB.89.184416} {\bibfield  {journal} {\bibinfo  {journal} {Phys.
  Rev. B}\ }\textbf {\bibinfo {volume} {89}},\ \bibinfo {pages} {184416}
  (\bibinfo {year} {2014})}\BibitemShut {NoStop}%
\bibitem [{\citenamefont {Zhang}\ \emph {et~al.}(2014)\citenamefont {Zhang},
  \citenamefont {Fritsch}, \citenamefont {Hao}, \citenamefont {Bagheri},
  \citenamefont {Gingras}, \citenamefont {Granroth}, \citenamefont
  {Jiramongkolchai}, \citenamefont {Cava},\ and\ \citenamefont
  {Gaulin}}]{Zhang14a}%
  \BibitemOpen
  \bibfield  {author} {\bibinfo {author} {\bibfnamefont {J.}~\bibnamefont
  {Zhang}}, \bibinfo {author} {\bibfnamefont {K.}~\bibnamefont {Fritsch}},
  \bibinfo {author} {\bibfnamefont {Z.}~\bibnamefont {Hao}}, \bibinfo {author}
  {\bibfnamefont {B.~V.}\ \bibnamefont {Bagheri}}, \bibinfo {author}
  {\bibfnamefont {M.~J.~P.}\ \bibnamefont {Gingras}}, \bibinfo {author}
  {\bibfnamefont {G.~E.}\ \bibnamefont {Granroth}}, \bibinfo {author}
  {\bibfnamefont {P.}~\bibnamefont {Jiramongkolchai}}, \bibinfo {author}
  {\bibfnamefont {R.~J.}\ \bibnamefont {Cava}}, \ and\ \bibinfo {author}
  {\bibfnamefont {B.~D.}\ \bibnamefont {Gaulin}},\ }\href {\doibase
  10.1103/PhysRevB.89.134410} {\bibfield  {journal} {\bibinfo  {journal} {Phys.
  Rev. B}\ }\textbf {\bibinfo {volume} {89}},\ \bibinfo {pages} {134410}
  (\bibinfo {year} {2014})}\BibitemShut {NoStop}%
\bibitem [{\citenamefont {Klekovkina}\ and\ \citenamefont
  {Malkin}(2014)}]{Klekovkina14a}%
  \BibitemOpen
  \bibfield  {author} {\bibinfo {author} {\bibfnamefont {V.}~\bibnamefont
  {Klekovkina}}\ and\ \bibinfo {author} {\bibfnamefont {B.}~\bibnamefont
  {Malkin}},\ }\href {\doibase 10.1134/S0030400X14060137} {\bibfield  {journal}
  {\bibinfo  {journal} {Optics and Spectroscopy}\ }\textbf {\bibinfo {volume}
  {116}},\ \bibinfo {pages} {849} (\bibinfo {year} {2014})}\BibitemShut
  {NoStop}%
\bibitem [{\citenamefont {Princep}\ \emph {et~al.}(2015)\citenamefont
  {Princep}, \citenamefont {Walker}, \citenamefont {Adroja}, \citenamefont
  {Prabhakaran},\ and\ \citenamefont {Boothroyd}}]{Princep15a}%
  \BibitemOpen
  \bibfield  {author} {\bibinfo {author} {\bibfnamefont {A.~J.}\ \bibnamefont
  {Princep}}, \bibinfo {author} {\bibfnamefont {H.~C.}\ \bibnamefont {Walker}},
  \bibinfo {author} {\bibfnamefont {D.~T.}\ \bibnamefont {Adroja}}, \bibinfo
  {author} {\bibfnamefont {D.}~\bibnamefont {Prabhakaran}}, \ and\ \bibinfo
  {author} {\bibfnamefont {A.~T.}\ \bibnamefont {Boothroyd}},\ }\href@noop {}
  {\bibfield  {journal} {\bibinfo  {journal} {ArXiv e-prints}\ } (\bibinfo
  {year} {2015})},\ \Eprint {http://arxiv.org/abs/1501.04927} {arXiv:1501.04927
  [cond-mat.str-el]} \BibitemShut {NoStop}%
\bibitem [{\citenamefont {Sazonov}\ \emph {et~al.}(2010)\citenamefont
  {Sazonov}, \citenamefont {Gukasov}, \citenamefont {Mirebeau}, \citenamefont
  {Cao}, \citenamefont {Bonville}, \citenamefont {Grenier},\ and\ \citenamefont
  {Dhalenne}}]{Sazonov10a}%
  \BibitemOpen
  \bibfield  {author} {\bibinfo {author} {\bibfnamefont {A.~P.}\ \bibnamefont
  {Sazonov}}, \bibinfo {author} {\bibfnamefont {A.}~\bibnamefont {Gukasov}},
  \bibinfo {author} {\bibfnamefont {I.}~\bibnamefont {Mirebeau}}, \bibinfo
  {author} {\bibfnamefont {H.}~\bibnamefont {Cao}}, \bibinfo {author}
  {\bibfnamefont {P.}~\bibnamefont {Bonville}}, \bibinfo {author}
  {\bibfnamefont {B.}~\bibnamefont {Grenier}}, \ and\ \bibinfo {author}
  {\bibfnamefont {G.}~\bibnamefont {Dhalenne}},\ }\href {\doibase
  10.1103/PhysRevB.82.174406} {\bibfield  {journal} {\bibinfo  {journal} {Phys.
  Rev. B}\ }\textbf {\bibinfo {volume} {82}},\ \bibinfo {pages} {174406}
  (\bibinfo {year} {2010})}\BibitemShut {NoStop}%
\bibitem [{\citenamefont {Enjalran}\ and\ \citenamefont
  {Gingras}(2004)}]{Enjalran04b}%
  \BibitemOpen
  \bibfield  {author} {\bibinfo {author} {\bibfnamefont {M.}~\bibnamefont
  {Enjalran}}\ and\ \bibinfo {author} {\bibfnamefont {M.~J.~P.}\ \bibnamefont
  {Gingras}},\ }\href {\doibase ARTN 174426} {\bibfield  {journal} {\bibinfo
  {journal} {Physical Review B}\ }\textbf {\bibinfo {volume} {70}},\ \bibinfo
  {pages} {174426} (\bibinfo {year} {2004})}\BibitemShut {NoStop}%
\bibitem [{\citenamefont {Mirebeau}\ \emph {et~al.}(2007)\citenamefont
  {Mirebeau}, \citenamefont {Bonville},\ and\ \citenamefont
  {Hennion}}]{Mirebeau07a}%
  \BibitemOpen
  \bibfield  {author} {\bibinfo {author} {\bibfnamefont {I.}~\bibnamefont
  {Mirebeau}}, \bibinfo {author} {\bibfnamefont {P.}~\bibnamefont {Bonville}},
  \ and\ \bibinfo {author} {\bibfnamefont {M.}~\bibnamefont {Hennion}},\ }\href
  {\doibase ARTN 184436} {\bibfield  {journal} {\bibinfo  {journal} {Physical
  Review B}\ }\textbf {\bibinfo {volume} {76}},\ \bibinfo {pages} {184436}
  (\bibinfo {year} {2007})}\BibitemShut {NoStop}%
\bibitem [{\citenamefont {Curnoe}(2007)}]{Curnoe07a}%
  \BibitemOpen
  \bibfield  {author} {\bibinfo {author} {\bibfnamefont {S.~H.}\ \bibnamefont
  {Curnoe}},\ }\href {\doibase ARTN 212404} {\bibfield  {journal} {\bibinfo
  {journal} {Physical Review B}\ }\textbf {\bibinfo {volume} {75}},\ \bibinfo
  {pages} {212404} (\bibinfo {year} {2007})}\BibitemShut {NoStop}%
\bibitem [{\citenamefont {Curnoe}(2008)}]{Curnoe08a}%
  \BibitemOpen
  \bibfield  {author} {\bibinfo {author} {\bibfnamefont {S.~H.}\ \bibnamefont
  {Curnoe}},\ }\href {\doibase ARTN 094418} {\bibfield  {journal} {\bibinfo
  {journal} {Physical Review B}\ }\textbf {\bibinfo {volume} {78}},\ \bibinfo
  {pages} {094418} (\bibinfo {year} {2008})}\BibitemShut {NoStop}%
\bibitem [{\citenamefont {McClarty}\ \emph {et~al.}(2014)\citenamefont
  {McClarty}, \citenamefont {Sikora}, \citenamefont {Moessner}, \citenamefont
  {Penc}, \citenamefont {Pollmann},\ and\ \citenamefont
  {Shannon}}]{Mcclarty14a}%
  \BibitemOpen
  \bibfield  {author} {\bibinfo {author} {\bibfnamefont {P.}~\bibnamefont
  {McClarty}}, \bibinfo {author} {\bibfnamefont {O.}~\bibnamefont {Sikora}},
  \bibinfo {author} {\bibfnamefont {R.}~\bibnamefont {Moessner}}, \bibinfo
  {author} {\bibfnamefont {K.}~\bibnamefont {Penc}}, \bibinfo {author}
  {\bibfnamefont {F.}~\bibnamefont {Pollmann}}, \ and\ \bibinfo {author}
  {\bibfnamefont {N.}~\bibnamefont {Shannon}},\ }\href@noop {} {\bibfield
  {journal} {\bibinfo  {journal} {ArXiv e-prints}\ } (\bibinfo {year}
  {2014})},\ \Eprint {http://arxiv.org/abs/1410.0451} {arXiv:1410.0451
  [cond-mat.str-el]} \BibitemShut {NoStop}%
\bibitem [{\citenamefont {Molavian}\ \emph {et~al.}(2007)\citenamefont
  {Molavian}, \citenamefont {Gingras},\ and\ \citenamefont
  {Canals}}]{Molavian07a}%
  \BibitemOpen
  \bibfield  {author} {\bibinfo {author} {\bibfnamefont {H.~R.}\ \bibnamefont
  {Molavian}}, \bibinfo {author} {\bibfnamefont {M.~J.~P.}\ \bibnamefont
  {Gingras}}, \ and\ \bibinfo {author} {\bibfnamefont {B.}~\bibnamefont
  {Canals}},\ }\href {\doibase ARTN 157204} {\bibfield  {journal} {\bibinfo
  {journal} {Physical Review Letters}\ }\textbf {\bibinfo {volume} {98}},\
  \bibinfo {pages} {157204} (\bibinfo {year} {2007})}\BibitemShut {NoStop}%
\bibitem [{\citenamefont {Molavian}\ \emph {et~al.}(2009)\citenamefont
  {Molavian}, \citenamefont {McClarty},\ and\ \citenamefont
  {Gingras}}]{Molavian09a}%
  \BibitemOpen
  \bibfield  {author} {\bibinfo {author} {\bibfnamefont {H.~R.}\ \bibnamefont
  {Molavian}}, \bibinfo {author} {\bibfnamefont {P.~A.}\ \bibnamefont
  {McClarty}}, \ and\ \bibinfo {author} {\bibfnamefont {M.~J.~P.}\ \bibnamefont
  {Gingras}},\ }\href@noop {} {\bibfield  {journal} {\bibinfo  {journal} {ArXiv
  e-prints}\ } (\bibinfo {year} {2009})},\ \Eprint
  {http://arxiv.org/abs/0912.2957} {arXiv:0912.2957 [cond-mat.stat-mech]}
  \BibitemShut {NoStop}%
\bibitem [{\citenamefont {Mirebeau}\ \emph {et~al.}(2002)\citenamefont
  {Mirebeau}, \citenamefont {Goncharenko}, \citenamefont {Cadavez-Pares},
  \citenamefont {Bramwell}, \citenamefont {Gingras},\ and\ \citenamefont
  {Gardner}}]{Mirebeau02a}%
  \BibitemOpen
  \bibfield  {author} {\bibinfo {author} {\bibfnamefont {I.}~\bibnamefont
  {Mirebeau}}, \bibinfo {author} {\bibfnamefont {I.~N.}\ \bibnamefont
  {Goncharenko}}, \bibinfo {author} {\bibfnamefont {P.}~\bibnamefont
  {Cadavez-Pares}}, \bibinfo {author} {\bibfnamefont {S.~T.}\ \bibnamefont
  {Bramwell}}, \bibinfo {author} {\bibfnamefont {M.~J.~P.}\ \bibnamefont
  {Gingras}}, \ and\ \bibinfo {author} {\bibfnamefont {J.~S.}\ \bibnamefont
  {Gardner}},\ }\href {\doibase DOI 10.1038/nature01157} {\bibfield  {journal}
  {\bibinfo  {journal} {Nature}\ }\textbf {\bibinfo {volume} {420}},\ \bibinfo
  {pages} {54} (\bibinfo {year} {2002})}\BibitemShut {NoStop}%
\bibitem [{\citenamefont {Taguchi}\ \emph {et~al.}(2001)\citenamefont
  {Taguchi}, \citenamefont {Oohara}, \citenamefont {Yoshizawa}, \citenamefont
  {Nagaosa},\ and\ \citenamefont {Tokura}}]{Taguchi01a}%
  \BibitemOpen
  \bibfield  {author} {\bibinfo {author} {\bibfnamefont {Y.}~\bibnamefont
  {Taguchi}}, \bibinfo {author} {\bibfnamefont {Y.}~\bibnamefont {Oohara}},
  \bibinfo {author} {\bibfnamefont {H.}~\bibnamefont {Yoshizawa}}, \bibinfo
  {author} {\bibfnamefont {N.}~\bibnamefont {Nagaosa}}, \ and\ \bibinfo
  {author} {\bibfnamefont {Y.}~\bibnamefont {Tokura}},\ }\href {\doibase
  10.1126/science.1058161} {\bibfield  {journal} {\bibinfo  {journal}
  {Science}\ }\textbf {\bibinfo {volume} {291}},\ \bibinfo {pages} {2573}
  (\bibinfo {year} {2001})}\BibitemShut {NoStop}%
\bibitem [{\citenamefont {Machida}\ \emph {et~al.}(2013)\citenamefont
  {Machida}, \citenamefont {Nakatsuji}, \citenamefont {Onoda}, \citenamefont
  {Tayama},\ and\ \citenamefont {Sakakibara}}]{Machida10a}%
  \BibitemOpen
  \bibfield  {author} {\bibinfo {author} {\bibfnamefont {Y.}~\bibnamefont
  {Machida}}, \bibinfo {author} {\bibfnamefont {S.}~\bibnamefont {Nakatsuji}},
  \bibinfo {author} {\bibfnamefont {S.}~\bibnamefont {Onoda}}, \bibinfo
  {author} {\bibfnamefont {T.}~\bibnamefont {Tayama}}, \ and\ \bibinfo {author}
  {\bibfnamefont {T.}~\bibnamefont {Sakakibara}},\ }\href {\doibase
  10.1038/nature08680} {\bibfield  {journal} {\bibinfo  {journal} {Nature}\
  }\textbf {\bibinfo {volume} {463}},\ \bibinfo {pages} {210} (\bibinfo {year}
  {2013})}\BibitemShut {NoStop}%
\bibitem [{\citenamefont {Nakatsuji}\ \emph {et~al.}(2006)\citenamefont
  {Nakatsuji}, \citenamefont {Machida}, \citenamefont {Maeno}, \citenamefont
  {Tayama}, \citenamefont {Sakakibara}, \citenamefont {van Duijn},
  \citenamefont {Balicas}, \citenamefont {Millican}, \citenamefont {Macaluso},\
  and\ \citenamefont {Chan}}]{Nakatsuji06a}%
  \BibitemOpen
  \bibfield  {author} {\bibinfo {author} {\bibfnamefont {S.}~\bibnamefont
  {Nakatsuji}}, \bibinfo {author} {\bibfnamefont {Y.}~\bibnamefont {Machida}},
  \bibinfo {author} {\bibfnamefont {Y.}~\bibnamefont {Maeno}}, \bibinfo
  {author} {\bibfnamefont {T.}~\bibnamefont {Tayama}}, \bibinfo {author}
  {\bibfnamefont {T.}~\bibnamefont {Sakakibara}}, \bibinfo {author}
  {\bibfnamefont {J.}~\bibnamefont {van Duijn}}, \bibinfo {author}
  {\bibfnamefont {L.}~\bibnamefont {Balicas}}, \bibinfo {author} {\bibfnamefont
  {J.~N.}\ \bibnamefont {Millican}}, \bibinfo {author} {\bibfnamefont {R.~T.}\
  \bibnamefont {Macaluso}}, \ and\ \bibinfo {author} {\bibfnamefont {J.~Y.}\
  \bibnamefont {Chan}},\ }\href {\doibase 10.1103/PhysRevLett.96.087204}
  {\bibfield  {journal} {\bibinfo  {journal} {Phys. Rev. Lett.}\ }\textbf
  {\bibinfo {volume} {96}},\ \bibinfo {pages} {087204} (\bibinfo {year}
  {2006})}\BibitemShut {NoStop}%
\bibitem [{\citenamefont {Sakata}\ \emph {et~al.}(2011)\citenamefont {Sakata},
  \citenamefont {Kagayama}, \citenamefont {Shimizu}, \citenamefont {Matsuhira},
  \citenamefont {Takagi}, \citenamefont {Wakeshima},\ and\ \citenamefont
  {Hinatsu}}]{Sakata11a}%
  \BibitemOpen
  \bibfield  {author} {\bibinfo {author} {\bibfnamefont {M.}~\bibnamefont
  {Sakata}}, \bibinfo {author} {\bibfnamefont {T.}~\bibnamefont {Kagayama}},
  \bibinfo {author} {\bibfnamefont {K.}~\bibnamefont {Shimizu}}, \bibinfo
  {author} {\bibfnamefont {K.}~\bibnamefont {Matsuhira}}, \bibinfo {author}
  {\bibfnamefont {S.}~\bibnamefont {Takagi}}, \bibinfo {author} {\bibfnamefont
  {M.}~\bibnamefont {Wakeshima}}, \ and\ \bibinfo {author} {\bibfnamefont
  {Y.}~\bibnamefont {Hinatsu}},\ }\href {\doibase 10.1103/PhysRevB.83.041102}
  {\bibfield  {journal} {\bibinfo  {journal} {Phys. Rev. B}\ }\textbf {\bibinfo
  {volume} {83}},\ \bibinfo {pages} {041102} (\bibinfo {year}
  {2011})}\BibitemShut {NoStop}%
\bibitem [{\citenamefont {Udagawa}\ \emph {et~al.}(2012)\citenamefont
  {Udagawa}, \citenamefont {Ishizuka},\ and\ \citenamefont
  {Motome}}]{Udagawa12a}%
  \BibitemOpen
  \bibfield  {author} {\bibinfo {author} {\bibfnamefont {M.}~\bibnamefont
  {Udagawa}}, \bibinfo {author} {\bibfnamefont {H.}~\bibnamefont {Ishizuka}}, \
  and\ \bibinfo {author} {\bibfnamefont {Y.}~\bibnamefont {Motome}},\ }\href
  {\doibase 10.1103/PhysRevLett.108.066406} {\bibfield  {journal} {\bibinfo
  {journal} {Phys. Rev. Lett.}\ }\textbf {\bibinfo {volume} {108}},\ \bibinfo
  {pages} {066406} (\bibinfo {year} {2012})}\BibitemShut {NoStop}%
\bibitem [{\citenamefont {Chern}\ \emph {et~al.}(2013)\citenamefont {Chern},
  \citenamefont {Maiti}, \citenamefont {Fernandes},\ and\ \citenamefont
  {W\"olfle}}]{Chern13a}%
  \BibitemOpen
  \bibfield  {author} {\bibinfo {author} {\bibfnamefont {G.-W.}\ \bibnamefont
  {Chern}}, \bibinfo {author} {\bibfnamefont {S.}~\bibnamefont {Maiti}},
  \bibinfo {author} {\bibfnamefont {R.~M.}\ \bibnamefont {Fernandes}}, \ and\
  \bibinfo {author} {\bibfnamefont {P.}~\bibnamefont {W\"olfle}},\ }\href
  {\doibase 10.1103/PhysRevLett.110.146602} {\bibfield  {journal} {\bibinfo
  {journal} {Phys. Rev. Lett.}\ }\textbf {\bibinfo {volume} {110}},\ \bibinfo
  {pages} {146602} (\bibinfo {year} {2013})}\BibitemShut {NoStop}%
\bibitem [{\citenamefont {Savary}\ \emph {et~al.}(2014)\citenamefont {Savary},
  \citenamefont {Moon},\ and\ \citenamefont {Balents}}]{Savary14a}%
  \BibitemOpen
  \bibfield  {author} {\bibinfo {author} {\bibfnamefont {L.}~\bibnamefont
  {Savary}}, \bibinfo {author} {\bibfnamefont {E.-G.}\ \bibnamefont {Moon}}, \
  and\ \bibinfo {author} {\bibfnamefont {L.}~\bibnamefont {Balents}},\ }\href
  {\doibase 10.1103/PhysRevX.4.041027} {\bibfield  {journal} {\bibinfo
  {journal} {Phys. Rev. X}\ }\textbf {\bibinfo {volume} {4}},\ \bibinfo {pages}
  {041027} (\bibinfo {year} {2014})}\BibitemShut {NoStop}%
\bibitem [{\citenamefont {Sarkar}\ and\ \citenamefont
  {Mukhopadhyay}(2014)}]{Sarkar14a}%
  \BibitemOpen
  \bibfield  {author} {\bibinfo {author} {\bibfnamefont {A.}~\bibnamefont
  {Sarkar}}\ and\ \bibinfo {author} {\bibfnamefont {S.}~\bibnamefont
  {Mukhopadhyay}},\ }\href {\doibase 10.1103/PhysRevB.90.165129} {\bibfield
  {journal} {\bibinfo  {journal} {Phys. Rev. B}\ }\textbf {\bibinfo {volume}
  {90}},\ \bibinfo {pages} {165129} (\bibinfo {year} {2014})}\BibitemShut
  {NoStop}%
\bibitem [{\citenamefont {Katsufuji}\ and\ \citenamefont
  {Takagi}(2004)}]{Katsufuji04a}%
  \BibitemOpen
  \bibfield  {author} {\bibinfo {author} {\bibfnamefont {T.}~\bibnamefont
  {Katsufuji}}\ and\ \bibinfo {author} {\bibfnamefont {H.}~\bibnamefont
  {Takagi}},\ }\href {\doibase 10.1103/PhysRevB.69.064422} {\bibfield
  {journal} {\bibinfo  {journal} {Phys. Rev. B}\ }\textbf {\bibinfo {volume}
  {69}},\ \bibinfo {pages} {064422} (\bibinfo {year} {2004})}\BibitemShut
  {NoStop}%
\bibitem [{\citenamefont {Saito}\ \emph {et~al.}(2005)\citenamefont {Saito},
  \citenamefont {Higashinaka},\ and\ \citenamefont {Maeno}}]{Saito05a}%
  \BibitemOpen
  \bibfield  {author} {\bibinfo {author} {\bibfnamefont {M.}~\bibnamefont
  {Saito}}, \bibinfo {author} {\bibfnamefont {R.}~\bibnamefont {Higashinaka}},
  \ and\ \bibinfo {author} {\bibfnamefont {Y.}~\bibnamefont {Maeno}},\ }\href
  {\doibase ARTN 144422} {\bibfield  {journal} {\bibinfo  {journal} {Physical
  Review B}\ }\textbf {\bibinfo {volume} {72}},\ \bibinfo {pages} {144422}
  (\bibinfo {year} {2005})}\BibitemShut {NoStop}%
\bibitem [{\citenamefont {Liu}\ \emph {et~al.}(2013)\citenamefont {Liu},
  \citenamefont {Lin}, \citenamefont {Liu}, \citenamefont {Yan}, \citenamefont
  {Dong},\ and\ \citenamefont {Liu}}]{Liu13a}%
  \BibitemOpen
  \bibfield  {author} {\bibinfo {author} {\bibfnamefont {D.}~\bibnamefont
  {Liu}}, \bibinfo {author} {\bibfnamefont {L.}~\bibnamefont {Lin}}, \bibinfo
  {author} {\bibfnamefont {M.~F.}\ \bibnamefont {Liu}}, \bibinfo {author}
  {\bibfnamefont {Z.~B.}\ \bibnamefont {Yan}}, \bibinfo {author} {\bibfnamefont
  {S.}~\bibnamefont {Dong}}, \ and\ \bibinfo {author} {\bibfnamefont {J.~M.}\
  \bibnamefont {Liu}},\ }\href {\doibase 10.1063/1.4793704} {\bibfield
  {journal} {\bibinfo  {journal} {Journal of Applied Physics}\ }\textbf
  {\bibinfo {volume} {113}} (\bibinfo {year} {2013}),\
  10.1063/1.4793704}\BibitemShut {NoStop}%
\bibitem [{\citenamefont {Grams}\ \emph {et~al.}(2014)\citenamefont {Grams},
  \citenamefont {Valldor}, \citenamefont {Garst},\ and\ \citenamefont
  {Hemberger}}]{Grams14a}%
  \BibitemOpen
  \bibfield  {author} {\bibinfo {author} {\bibfnamefont {C.~P.}\ \bibnamefont
  {Grams}}, \bibinfo {author} {\bibfnamefont {M.}~\bibnamefont {Valldor}},
  \bibinfo {author} {\bibfnamefont {M.}~\bibnamefont {Garst}}, \ and\ \bibinfo
  {author} {\bibfnamefont {J.}~\bibnamefont {Hemberger}},\ }\href {\doibase
  10.1038/ncomms5853} {\bibfield  {journal} {\bibinfo  {journal} {Nature
  Communications}\ }\textbf {\bibinfo {volume} {5}} (\bibinfo {year} {2014}),\
  10.1038/ncomms5853}\BibitemShut {NoStop}%
\bibitem [{\citenamefont {Zhou}\ \emph {et~al.}(2012)\citenamefont {Zhou},
  \citenamefont {Cheng}, \citenamefont {Hallas}, \citenamefont {Wiebe},
  \citenamefont {Li}, \citenamefont {Balicas}, \citenamefont {Zhou},
  \citenamefont {Goodenough}, \citenamefont {Gardner},\ and\ \citenamefont
  {Choi}}]{Zhou12a}%
  \BibitemOpen
  \bibfield  {author} {\bibinfo {author} {\bibfnamefont {H.~D.}\ \bibnamefont
  {Zhou}}, \bibinfo {author} {\bibfnamefont {J.~G.}\ \bibnamefont {Cheng}},
  \bibinfo {author} {\bibfnamefont {A.~M.}\ \bibnamefont {Hallas}}, \bibinfo
  {author} {\bibfnamefont {C.~R.}\ \bibnamefont {Wiebe}}, \bibinfo {author}
  {\bibfnamefont {G.}~\bibnamefont {Li}}, \bibinfo {author} {\bibfnamefont
  {L.}~\bibnamefont {Balicas}}, \bibinfo {author} {\bibfnamefont {J.~S.}\
  \bibnamefont {Zhou}}, \bibinfo {author} {\bibfnamefont {J.~B.}\ \bibnamefont
  {Goodenough}}, \bibinfo {author} {\bibfnamefont {J.~S.}\ \bibnamefont
  {Gardner}}, \ and\ \bibinfo {author} {\bibfnamefont {E.~S.}\ \bibnamefont
  {Choi}},\ }\href {\doibase 10.1103/PhysRevLett.108.207206} {\bibfield
  {journal} {\bibinfo  {journal} {Phys. Rev. Lett.}\ }\textbf {\bibinfo
  {volume} {108}},\ \bibinfo {pages} {207206} (\bibinfo {year}
  {2012})}\BibitemShut {NoStop}%
\bibitem [{\citenamefont {Lin}\ \emph {et~al.}(2013)\citenamefont {Lin},
  \citenamefont {Zhao}, \citenamefont {Liu}, \citenamefont {Xie}, \citenamefont
  {Dong}, \citenamefont {Yan},\ and\ \citenamefont {Liu}}]{Lin13a}%
  \BibitemOpen
  \bibfield  {author} {\bibinfo {author} {\bibfnamefont {L.}~\bibnamefont
  {Lin}}, \bibinfo {author} {\bibfnamefont {Z.~Y.}\ \bibnamefont {Zhao}},
  \bibinfo {author} {\bibfnamefont {D.}~\bibnamefont {Liu}}, \bibinfo {author}
  {\bibfnamefont {Y.~L.}\ \bibnamefont {Xie}}, \bibinfo {author} {\bibfnamefont
  {S.}~\bibnamefont {Dong}}, \bibinfo {author} {\bibfnamefont {Z.~B.}\
  \bibnamefont {Yan}}, \ and\ \bibinfo {author} {\bibfnamefont {J.~M.}\
  \bibnamefont {Liu}},\ }\href {\doibase 10.1063/1.4794129} {\bibfield
  {journal} {\bibinfo  {journal} {Journal of Applied Physics}\ }\textbf
  {\bibinfo {volume} {113}} (\bibinfo {year} {2013}),\ 10.1063/1.4794129},\
  \bibinfo {note} {12th Joint MMM-Intermag Conference, Chicago, IL, JAN 14-18,
  2013}\BibitemShut {NoStop}%
\bibitem [{\citenamefont {Xu}\ \emph {et~al.}(2014)\citenamefont {Xu},
  \citenamefont {Liu}, \citenamefont {Lin}, \citenamefont {Liu}, \citenamefont
  {Yan},\ and\ \citenamefont {Liu}}]{Xu14a}%
  \BibitemOpen
  \bibfield  {author} {\bibinfo {author} {\bibfnamefont {Z.-C.}\ \bibnamefont
  {Xu}}, \bibinfo {author} {\bibfnamefont {M.-F.}\ \bibnamefont {Liu}},
  \bibinfo {author} {\bibfnamefont {L.}~\bibnamefont {Lin}}, \bibinfo {author}
  {\bibfnamefont {H.}~\bibnamefont {Liu}}, \bibinfo {author} {\bibfnamefont
  {Z.-B.}\ \bibnamefont {Yan}}, \ and\ \bibinfo {author} {\bibfnamefont
  {J.-M.}\ \bibnamefont {Liu}},\ }\href {\doibase 10.1007/s11467-013-0395-8}
  {\bibfield  {journal} {\bibinfo  {journal} {Frontiers of Physics}\ }\textbf
  {\bibinfo {volume} {9}},\ \bibinfo {eid} {82} (\bibinfo {year}
  {2014})}\BibitemShut {NoStop}%
\bibitem [{\citenamefont {Brooks-Bartlett}\ \emph {et~al.}(2014)\citenamefont
  {Brooks-Bartlett}, \citenamefont {Banks}, \citenamefont {Jaubert},
  \citenamefont {Harman-Clarke},\ and\ \citenamefont {Holdsworth}}]{Brooks14a}%
  \BibitemOpen
  \bibfield  {author} {\bibinfo {author} {\bibfnamefont {M.}~\bibnamefont
  {Brooks-Bartlett}}, \bibinfo {author} {\bibfnamefont {S.}~\bibnamefont
  {Banks}}, \bibinfo {author} {\bibfnamefont {L.}~\bibnamefont {Jaubert}},
  \bibinfo {author} {\bibfnamefont {A.}~\bibnamefont {Harman-Clarke}}, \ and\
  \bibinfo {author} {\bibfnamefont {P.}~\bibnamefont {Holdsworth}},\
  }\href@noop {} {\bibfield  {journal} {\bibinfo  {journal} {Phys. Rev. X}\
  }\textbf {\bibinfo {volume} {4}},\ \bibinfo {pages} {011007} (\bibinfo {year}
  {2014})}\BibitemShut {NoStop}%
\end{thebibliography}%

\end{document}